\documentclass{wileyvch}
\usepackage{graphicx}
\usepackage{wileyfnt}
\pdfoutput=1
\usepackage{makeidx}
\usepackage{bbm}
\usepackage{mathrsfs}
\usepackage{amsmath,amsfonts,amssymb}
\makeindex

\newcommand{\bbM}{\mathbb M}

\copyrightinfo
{}% title
{}% author/authors/editor/editors
{}% ISBN number

\begin{document}
%%% Book title and half title page

%%% Set up for Half Title and Title Page:

%% All commands optional except for 
%% title and either author or editor:

%\author{Radons, Rumpf, Schuster (Eds.)}
%\title{Nonlinear Dynamics of Nanosystems}
%\subtitle{}
%\editor{}
%\edition{}
%\collaborators{}

%% for example:
% \author{Dietrich Klemm}

% \title{Biochemical Principles and\\
% Mechanisms of Biosynthesis\\
% Biodegradation of Polymers}

% \subtitle{Nomenclature of\\
% Organic Chemistry}

% \editor{Markus Stoeppler}

% \edition{Third Edition}

% \collaborators{Gerard B\"ohm\\
% Uwe Diderichsen}

%%% Print Half title and Titlepage: Comment out if you don't want
%%% the title pages to print

%\halftitle

%\titlepage

%%% Frontmatter
% Frontmatter in this order (tableofcontents is necessary, others optional)

%%%%%%%%%

%\dedication{}

% for example:
%\dedication{for Hans Martin}

%%%%%%%%%
%\foreword{}

%\forewordauthor{}% city, date
{}% authors/editors

% for example:
% \foreword{A Personal Foreword}
% The dilemma of rapidly emerging fields is that reviews are often...
%
%\forewordauthor{Basel, March 2006}% city, date
%{Wolfgang Janke\\ Daniel A. Erlanson}% authors/editors

%%%%%%%%%
%\tableofcontents
%\listoffigures
%\listoftables

%%%%%%%%%
%\preface{}% Preface title, you choose title

%\prefaceauthor{}% city, date
{}% author/editors

% for example:
% \preface{Preface to the Third Edition}
% Here is preface to the current edition.
% \prefaceauthor{Basel, March 2006}{Wolfgang Janke\\
% Daniel A. Erlanson}

%%% End Frontmatter

%%%%%%%%
%% Optional Part

%\part{}% Part title

% for example:
% \part{Sample Part}

%%%%%%%%
%% Chapter title

%%    If you want to send an
%%    alternate version of chapter title to the TOC and running head, enter
%%    it in square brackets, ie

%% \chapter[version for TOC and running head] 
%% {version to be printed on page} 

%% You may enter a footnote in the chapter title

% For example:
% \chapter[Short version of Chapter Title]
% {Nuclear Angular Momentum $\alpha\beta\Gamma\Delta 123$
% and Magnetic Moment 
% \footnote{Text for footnote.}}

%%% Second chapter example:
%  This example has multiple lines, so a version without \\ s
%  should be sent to the running head and Table of Contents with
%  the [] argument.

%% If you use \markboth{}{} remember to enter \thechapter\hskip1em
%% to send the chapter number to the running head.

% \chapter[Trust and Reputation For Service-Oriented
% Environments Technologies For Building Business Intelligence
% and Consumer Confidence]
% {Trust and Reputation\\ For Service-Oriented
% Environments Technologies\\ For Building Business Intelligence\\
% and Consumer Confidence}
% \markboth{\thechapter\hskip1em Trust and Reputation
% for Service-Oriented Environments Technologies}
% {\thechapter\hskip1em Trust and Reputation
% for Service-Oriented Environments Technologies}

%%%%%%%%
%% Edited Book, Multiple Authors

\chapter{Casimir Forces and Geometry in Nanosystems}
\chapauthor{Thorsten Emig\footnote{Institut f\"ur Theoretische Physik, Universit\"at zu K\"oln, Z\"ulpicher Strasse 77, 50937 K\"oln, Germany;
Laboratoire de Physique Th\'eorique et Mod\`eles
  Statistiques, CNRS UMR 8626, B\^at.~100, Universit\'e Paris-Sud, 91405
  Orsay cedex, France}} 

%% If there is a chapter author, enter name here

%\chapauthor{Chapter Author\footnote{Chapter author footnote}}

% If \chapauthor{} is used, make first paragraph afterwards be not
% indented, using \noindent, for example:

\noindent

Casimir interactions, predicted in 1948
\cite{Casimir:1948a,Casimir:1948dh} between atoms and macroscopic
surfaces, and probed in a series of high precision experiments over
the past decade \cite{Lamoreaux:1997a,Mohideen:1998a,Bressi:2002a},
are particular important at micro-meter to nano-meter length scales
due to their strong power-law increase at short separations between
particles.  Therefore, in constructing and operating devices at these
length scales it is important to have an accurate understanding of
material, shape and geometry dependence of these forces.  In
particular, the observation of Casimir forces in devices on sub-micron
scales has generated great current interest in exploring the role of
these forces for the development and optimization of micro- and
nano-electromechanical systems
\cite{Buks:2001b,Chan:2001a,Chan:2001d}. These systems can serve as
on-chip fully integrated sensors and actuators with a growing number
of applications. It was pointed out that Casimir forces can make an
important contribution to the principal cause of malfunctions of these
devices in form of stiction that results in permanent adhesion of
nearby surface elements \cite{Buks:2001a}. This initiated interest in
repulsive Casimir forces by modifying material properties as well as
the geometry of the interacting components
\cite{Buks:2002a,Kenneth:2002b,Munday:2009xr}.

The study of fluctuation induced forces has a long history.  When
these forces result from fluctuations of charges and currents inside
particles or macroscopic objects, they are usually summarized under
the general term van der Waals forces \cite{Parsegian:2005eu}. This
interaction appears at the atomic scale in the guise of Keesom, Debye,
London, and Casimir-Polder forces. An important property of all this
interactions is their non-additivity: The total interaction of
macroscopic objects is in general not given by the sum of the
interactions between all pairs of particles that form the objects.
This inherent many-body character of the force leads to interesting
and often unexpected behavior but renders the study of these forces
also a difficult problem. Commonly used approximations as pairwise
additivity assumptions become unreliable for systems of condensed
atoms. The collective interaction of condensed macroscopic systems is
better formulated in terms of their dielectric properties.  Such a
formulation was established by Lifshitz for two parallel and planar,
infinitely extended dielectric surfaces \cite{Lifshitz:1956ud},
extending Casimir's original work for perfect metals. In practice, one
encounters objects of finite size with curved surfaces and/or edges,
like structured surfaces, spheres or cylinders.  Also, in small-scale
devices often more than two objects are at close separation and one
would like to know the collective effects resulting from
non-additivity. In this chapter we shall encounter a selection of
examples for interesting behavior of fluctuation forces that result
from shape and material properties that have been obtained from a
recently developed method that makes it possible to compute van der
Waals-Casimir interactions for arbitrary compact objects based on
their scattering properties for electromagnetic waves
\cite{Emig:2007os,Emig:2008qf,Emig:2008ee,Kenneth:2008lo}.

%%%%%%%%
% Section commands, alternate version in square bracket 
% can be sent to running head
% and table of contents. 
% \\ can be used to start new line in two line section head
% \footnote{} can be used
% \sectionauthor{} can be used for author of particular section
% \markright{} may be used After section head to send different
% version of section title to the running head.

\section{Casimir effect}
\label{sec:Casimir_effect}

The Casimir effect is the attraction between two uncharged, parallel
and perfectly conducting plates \cite{Casimir:1948dh}.  For this
simple geometry the interaction can be obtained directly from the
plate induced {\it change} of the energies of the quantum mechanical
harmonics oscillators associated with the normal modes of the
electromagnetic field. The derivation given here follows closely the
one originally presented by Casimir. Consider two parallel and planar
surfaces of size $L \times L$ and separation $d$.  We assume that the
system is at zero temperature so that the interaction is given by the
ground state energies of harmonic oscillators. When we are interested
in the pressure (force per plate area $L^2$) between large plates with
$L\gg d$, we can ignore edge effects $\sim L$ and allow for a
continuum of wave vectors parallel to the plates. For a perfect
conductor the tangential electric field has to vanish at the surface
and the normal modes correspond to the allowed wave vectors ${\bf k}=({\bf
  k}_\|,\pi n/d)$ where ${\bf k}_\|$ is the two-dimensional wave
vector parallel to the plates. The linear dispersion of photons yields
the eigenfrequencies $\omega_{n k_\|}=c\sqrt{k_\|^2+(\pi n
  /d)^2}$ so that the ground state energy becomes
\begin{equation}
  \label{eq:plates_gs_energy}
  E = \frac{\hbar}{2}\, {\sum_{n=0}^{\infty}}' \left(\frac{L}{2\pi}\right)^2
 \int d^2 k_\| \, 2 \omega_{n k_\|} \, , 
\end{equation}
where we have included a factor of $2$ since for each mode with $n\neq
0$ exist two polarizations. The primed summation assigns a weight of
$1/2$ to the term for $n=0$. Obviously, the expression of
Eq.~(\ref{eq:plates_gs_energy}) is divergent. This is a consequence of
the assumption that the surfaces behave as a perfect conductor for
arbitrarily high frequencies. In practice, as pointed out by Casimir,
for very high frequencies (X-rays, e.g.) the plates are hardly an
obstacle for electromagnetic waves and therefore the ground state
energy of these modes will not be changed by the presence of the
plates. We implement this observation by introducing a cut-off
function $\chi(z)$ that is regular at $z=0$ with $\chi(0)=1$ and
vanishes, along with all its derivatives, for $z\to\infty$
sufficiently fast. After a change of variables,
$\omega=c\sqrt{k_\|^2+(\pi n/d)^2}$ with $c^2 k_\| dk_\| = \omega
d\omega$, we obtain the finite expression for the energy
\begin{equation}
  \label{eq:plates_energy_reg}
  E = \frac{\hbar L^2}{2\pi c^2} {\sum_{n=0}^{\infty}}' f(n) \quad
{\rm with} \quad f(n)=\int_{\pi nc/d}^\infty
\, \omega^2 \chi(\omega/\omega_c) d\omega 
\end{equation}
where $\omega_c$ is a cut-off frequency. As mentioned before, we are
interested in the change of the energy due the presence of the plates.
Let us imagine that we increase the separation $d$ between the plates
to infinity, thus creating empty space. When we subtract the energy of
the latter configuration from the total energy of
Eq.~(\ref{eq:plates_energy_reg}), we obtain the change in energy that
is the relevant interaction potential between the plates.  When the
separation $d$ tends to infinity, the sum in
Eq.~(\ref{eq:plates_energy_reg}) can be replaced by an integral,
yielding after the substitution $\Omega=\pi n c /d$ the energy
\begin{equation}
  \label{eq:E_infty}
  E_\infty = \frac{\hbar}{2\pi^2 c^3} \, L^2 d \, \int_0^\infty 
d \Omega \tilde f(\Omega) \quad {\rm with} \quad 
\tilde f(\Omega)=\int_\Omega^\infty \omega^2 \chi(\omega/\omega_c) d\omega \, .
\end{equation}
As expected, the energy $E_\infty$ is proportional to the volume $L^2
d$ of empty space and to a cut-off dependent factor that is given by the
integrals of Eq.~(\ref{eq:E_infty}).  This factor describes the
self-energy of the bounding surfaces. It is infinite for perfect
conductors which correspond to $\omega_c\to \infty$. For a non-ideal
conductor or any other material this factor is finite but depends on
material properties like the plasma wavelength for a metal. Now
we compute the change in energy when the plates are moved in
from infinity,
\begin{equation}
  \label{eq:delta_e}
  \Delta E = E - E_\infty = \frac{\hbar L^2}{2\pi c^2} \left[
 {\sum_{n=0}^{\infty}}' f(n) - \int_0^\infty dn f(n) \right] \, .
\end{equation}
The difference between the sum and the integral is given by the
Euler-Maclaurin formula ${\sum_{n=0}^{\infty}}' f(n) - \int_0^\infty
dn f(n)= -\frac{1}{12} f'(0) +\frac{1}{6!} f'''(0) + {\cal
  O}(f^v(0))$. This series of derivatives of odd order can be
truncated in the limit of perfect conductors $\omega_c\to\infty$ since
$f'(0)=0$, $f'''(0)=-2(\pi c/d)^3$ and $f^{(\nu)}(0)\sim
(c/d)^3 (c/d\omega_c)^{\nu-3}$. The Casimir potential hence
becomes
\begin{equation}
  \label{eq:Casimir_E_plates}
  \Delta E = - \frac{\pi^2}{720}\,\frac{\hbar c}{d^3}\, L^2  + {\cal O}(\omega_c^{-2})\, ,
\end{equation}
and the pressure for perfect metal plates is
\begin{equation}
  \label{eq:plates_pressure}
  \frac{F}{L^2} = - \frac{\pi^2}{240}\,\frac{\hbar c}{d^4} \, .
\end{equation}
The interesting fact is that the amplitude of the interaction is
{\it universal}, i.e., independent of the cut-off that can be viewed as a
simplified description of a real metal. This implies that for any pair
of surfaces with metallic response in the limit of small frequencies
$\omega\to 0$ the interaction at asymptotically large separations is
described by the potential of Eq.~(\ref{eq:Casimir_E_plates}). At a
separation of $d=100$nm Eq.~(\ref{eq:plates_pressure}) yields a
pressure of  $1.28 \cdot 10^{-4}\,$atm or $13.00\,$Pa.

% For example:
% \section[Section Title]
% {Section Title\footnote{Here is a section footnote}}
% \sectionauthor{Wally Jumblatt, Stephen Queen\footnote{Section Author}}

%%%%%%%%
% Subsection commands, alternate version in square bracket 
% can be sent to running head
% and table of contents. 
% \\ can be used to start new line in two line section head
% \footnote{} can be used
% \sectionauthor{} can be used for author of particular subsection

% For example:
% \subsection{Subsection Title\footnote{Footnote in Subsection}}
% \sectionauthor{Glen Greenwald}

\section{Dependence on Shape and Geometry}

Casimir interactions result from a modification of the fluctuation
spectrum of the electromagnetic field due to boundaries or coupling to
matter. This suggest that these interactions strongly depend on the
shape of the interacting objects and geometry, i.e., relative
position and orientation. The most commonly encountered geometry is a
sphere-plate setup that was used in the first high-precision tests of
the Casimir effect \cite{Lamoreaux:1997a,Mohideen:1998a}. This geometry has
been successfully used ever since in most of the experimental
studies of Casimir forces between metallic surfaces
\cite{Roy:1999b,Ederth:2000a,Chan:2001a,Chen:2002a,Chen:2006c,Decca:2007a,Chen:2007b,Munday:2007a}.
In order to keep the deviations from two parallel plates sufficiently
small, a sphere with a radius much larger than the surface distance
has been used.  The effect of curvature has been accounted for by the
``proximity force approximation'' (PFA) \cite{Parsegian:2005eu}. This scheme
is assumed to describe the interaction for sufficiently small ratios
of radius of curvature to distance. However, this an uncontrolled
assumption since PFA becomes exact only for infinitesimal separations,
and corrections to PFA are generally unknown.

At the other extreme, the interaction between a planar surface and
an object that is either very small or at an asymptotically large distance
is governed by the Casimir-Polder potential that was derived for the
case of an atom and a perfectly conducting plane
\cite{Casimir:1948a}. There have been attempts to go beyond the two
extreme limits of asymptotically large and small separations by
measuring the Casimir force between a sphere and a plane over a larger
range of ratios of sphere radius to distance \cite{Krause:2007a}.

Until very recently, no practical tools were available to compute the
electromagnetic Casimir interaction between objects of arbitrary shape
at all distances, including the important sphere-plate geometry.
Progress in understanding the geometry dependence of fluctuation
forces was hampered by the lack of methods that are applicable over a
wide range of separations. Unlike the case of parallel plates, the
eigenvalues of the Helmholtz equation in more complicated geometries
are in general not known and a summation over normal modes, as in the
original Casimir calculation of Section \ref{sec:Casimir_effect}, is
not practical.  Conceptually, the effects of geometry and shape are
difficult to study due to the non-additivity of fluctuation forces.

For some decades, there has been considerable interest in the theory
of Casimir forces between objects with curved surfaces. Two types of
approaches have been pursued. Attempts to compute the force explicitly
in particular geometries and efforts to develop a general framework
which yields the interaction in terms of characteristics of the
objects like polarizability or curvature. Within the second type of
approach, Balian and Duplantier studied the electromagnetic Casimir
interaction between compact and perfect metallic shapes in terms of a
multiple reflection expansion and derived also explicit results to
leading order at asymptotically large separations
\cite{Balian:1977a,BALIAN:1978a}.  For parallel and partially
transmitting plates a connection to scattering theory has been
established which yields the Casimir interaction of the plates as a
determinant of a diagonal matrix of reflection amplitudes
\cite{JAEKEL:1991a}. For non-planar, deformed plates, a general
representation of the Casimir energy as a functional determinant of a
matrix that describes reflections at the surfaces and free propagation
between them has been developed in Ref.~\cite{Emig:2004a}. Later on, an
equivalent representation has been applied to perturbative
computations in the case of rough and corrugated plates with finite
conductivity \cite{Lambrecht:2006a,Neto:2005a,Neto:2005c}.

Functional determinant formulas have been used also for open
geometries that do not fall into the class of parallel plates with
deformations.  For the Casimir interaction between planar plates and
cylinders, a partial wave expansion of the functional determinant has
been employed \cite{Emig:2006a,Bordag:2006b}.  More recently, a new
method based on a multipole expansion of fluctuating currents inside
the objects has been developed
\cite{Emig:2007os,Emig:2008qf,Emig:2008ee}. This method allows for
accurate and efficient calculations of Casimir forces and torques
between compact objects of arbitrary shape and material composition in
terms of the scattering matrices of the {\it individual} objects.  A
similar scattering approach has been developed in Ref.
\cite{Kenneth:2008lo}.

In this Section three examples for the strong geometry dependence of
Casimir forces will be made explicit. First, an overview on forces
between {\it deformed and structured surfaces} will be given. The
interactions are obtained from both a perturbative and numerical
evaluation of a functional determinant representation of Casimir
interactions between ideal metal surfaces. As a second example we
describe the effects that occur in the interaction of {\it
  one-dimensional structures as cylinders and wires} and related
non-additivity phenomena for more than two objects. Finally, as an
example for the interaction between compact objects, the Casimir force
between {\it metallic spheres} is presented for the full range of
separations, covering the crossover from the asymptotic Casimir-Polder
law to proximity approximations.  The analysis of the last two
examples is based on a scattering approach.

\subsection{Deformed surfaces}

The dependence of the Casimir force on shape and material properties
offers the opportunity to manipulate this interaction in a controlled
way, e.g., by imprinting patterns on the interacting surfaces. It has
been shown that a promising route to this end is via modifications of
the parallel plate geometry
\cite{Golestanian:1998a,Emig:2001a,Emig:2003a}. The corrections due to
deformations, such as sinusoidal corrugations, of the metal plates can
be significant. In searching for non-trivial shape dependences, Roy
and Mohideen \cite{Roy:1999a} measured the force between a sphere with
large radius and a sinusoidally corrugated plate with amplitude
$a\approx 60$nm and wavelength $\lambda\approx 1.1\mu$m.  Over the
range of separations $H\approx 0.1 - 0.9 \, \mu\mbox{m}$, the observed
force showed clear deviations from the dependence expected on the
basis of decomposing the Casimir force to a sum of pairwise
contributions (in effect, an average over the variations in
separations). Motivated by this experiment, the effect of corrugations
on the Casimir force between surfaces has been studied without using
pairwise additivity approximations. The analysis is based on a path
integral quantization of the fluctuating field with appropriate
boundary conditions which leads to a functional determinant
representation of the Casimir energy \cite{Golestanian:1998a} which
can be evaluated perturbatively for a small deformation amplitude
\cite{Emig:2001a,Emig:2003a} or numerically
for general amplitudes \cite{Emig:2003b,Buscher:2004a,Buscher:2005a}. In recent experiments
\cite{Chan:2008a} the Casimir force between a gold sphere and a
silicon surface with an array of nanoscale, rectangular corrugations
has been measured and the results were found to be consistent with the
theory based on a numerical evaluation of functional determinants for
ideal metals (see below). Qualitative agreement can be expected only
when material properties are taken into account in addition to shape.

\begin{figure}[h]
\includegraphics[width=7cm]{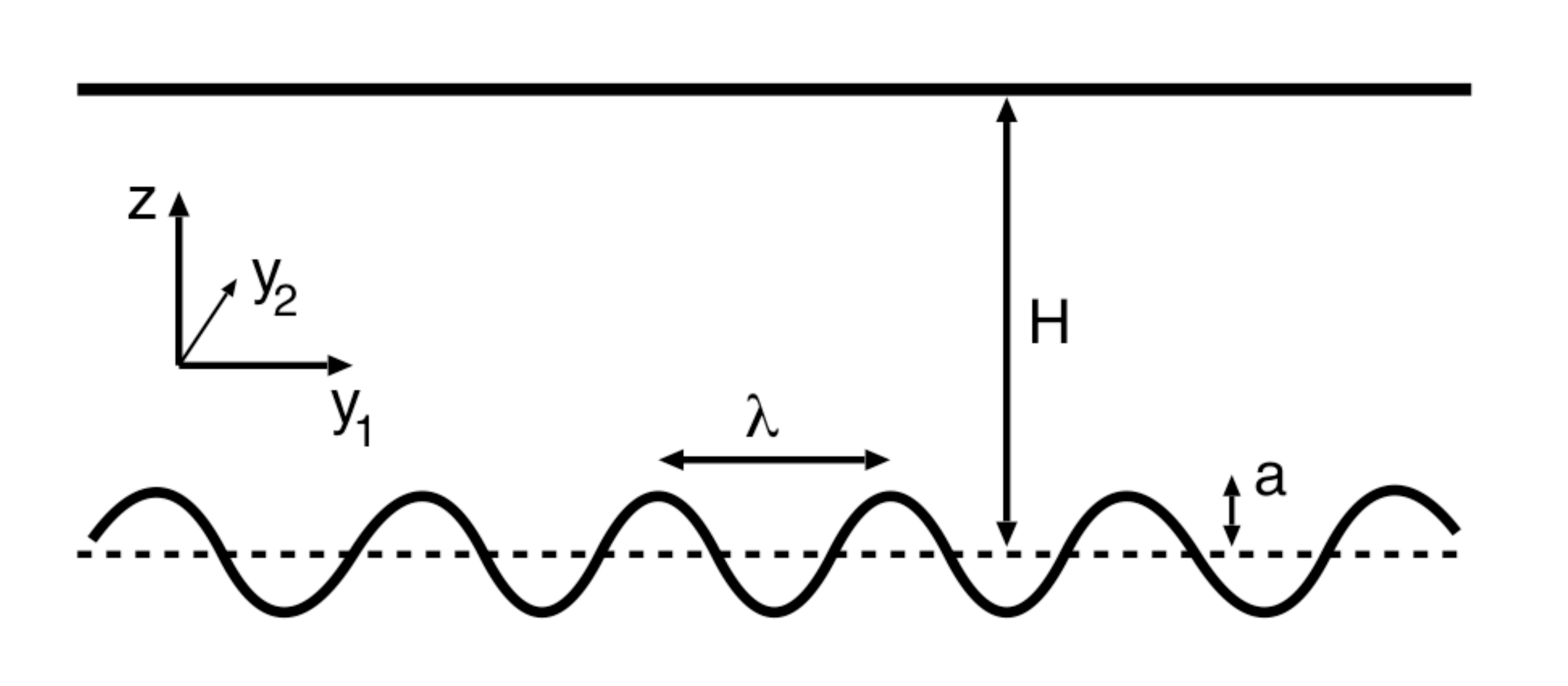}
\caption{\label{Fig:corr-flat} Configuration of a flat and a
  corrugated plate at mean separation $H$.}
\end{figure}

While we will be interested in the interaction between flat and
corrugated surfaces as depicted in Figs.~\ref{Fig:corr-flat},
\ref{Fig:corr-corr}, first of all we consider two surfaces with
arbitrary {\it uniaxial} deformations without overhangs so that their
profiles can be described by height functions $h_\alpha(y_1)$
($\alpha=1,2$ for the two surfaces), with $\int d y_1 \,
h_{\alpha}(y_1) = 0$.  It is further assumed that the surfaces are
perfectly conducting and infinitely extended along the plane spanned
by ${\bf y}_{\parallel} = (y_1, y_2)$.  As explained before for two
planar plates, the Casimir energy at zero temperature corresponds to
the difference of the ground state energies of the quantized
electromagnetic field for plates at distance $H$ and at $H \to
\infty$, respectively.  To obtain this energy, we employ a path
integral quantization method. For general, non-uniaxial deformations
or objects of more general shape, it is necessary to consider the
action for the electromagnetic field since the two polarizations (TM
for transverse magnetic waves and TE for transverse electric waves)
are coupled.  However, for the uniaxial deformations under
consideration here, we can develop a simpler quantization scheme, by a
similar reasoning as used in the context of waveguides with constant
cross-sectional shape \cite{Emig:2001a}. In this case, the two
polarizations are {\it independent} modes which do not couple under
scattering between the surfaces.  For TM waves all field components
are then fully specified by a scalar field corresponding to the
electric field along the invariant direction,
\begin{equation} \label{TM}
\Phi_{\rm TM}(t, y_1, y_2, z) = E_2(t, y_1, y_2, z),
\end{equation}
with the Dirichlet boundary condition
$\Phi_{\rm TM}|_{S_\alpha} = 0$ on each surface $S_\alpha$.
The TE waves are analogously described by the scalar field
\begin{equation} \label{TE}
\Phi_{\rm TE}(t, y_1, y_2, z) = B_2(t, y_1, y_2, z),
\end{equation}
with the Neumann boundary condition $\partial_n
\Phi_{\rm TE}|_{S_\alpha} = 0$, where $\partial_n$ is the
normal derivative of the surface $S_\alpha$ pointing into the
space between the two plates.  After a Wick rotation to the
imaginary time variable $X^0 = i c t$, both fields
$\Phi_{\rm TM}$ and $\Phi_{\rm TE}$ can be quantized using
the Euclidean action
\begin{equation} \label{e-action}
S\{\Phi\} \, = \, \frac{1}{2} \int d^4 X \, (\nabla
\Phi)^2 \, \, .
\end{equation}

In order to obtain the change in the ground state energy that is
associated with the presence of the plates, we now consider the
partition functions ${\cal Z}_{\rm D}$ and ${\cal Z}_{\rm N}$ for the
scalar field Euclidean action both with Dirichlet (D) and Neumann (N)
boundary conditions at the surfaces. We implement the boundary
conditions on the surfaces $S_\alpha$ using delta functions, which
leads to the partition functions
%\begin{subequations}
\begin{eqnarray}
{\cal Z}_{\rm D} \, & = & \, \frac{1}{{\cal Z}_0} \int {\cal
D}\Phi \prod_{\alpha=1}^2 \prod_{X_{\alpha}}
\delta[\Phi(X_{\alpha})]
\exp(-S\{\Phi\}/\hbar)\, , \label{ZD} \\[2mm]
{\cal Z}_{\rm N} \, & = & \, \frac{1}{{\cal Z}_0} \int {\cal
D}\Phi \prod_{\alpha=1}^2 \prod_{X_{\alpha}} \delta[\partial_n
\Phi(X_{\alpha})] \exp(-S\{\Phi\}/\hbar) \label{ZN} \,
\, ,
\end{eqnarray}
%\end{subequations}
where ${\cal Z}_0$ is the partition function of the space without
plates. Here $X_1({\bf y}) = [{\bf y}, h_1(y_1)]$ and $X_2({\bf y}) =
[{\bf y}, H + h_2(y_1)]$, where ${\bf y} = (y_0, y_1, y_2) = (y_0,
{\bf y}_{\parallel})$, and $y_0 = i c t$, is a parametrization of the
plates in 4D Euclidean space.  The Casimir energy ${\cal E}$ per unit
area (at zero temperature) that results from moving the plates in from
infinity is obtained from the partition function as
\begin{equation} \label{cale}
{\cal E}(H)  = E(H)  - \lim_{H \to \infty} E(H)\, ,
\end{equation}
with
\begin{equation} \label{e}
E(H) = - \frac{\hbar c}{A L} \left[ \ln {\cal Z}_{\rm D} + \ln
{\cal Z}_{\rm N} \right],
\end{equation}
where $A$ is the surface area of the plates and the limit where the
overall Euclidean length in time direction, $L$, tends to infinity is
implicitly assumed. The partition functions can be expressed as
functional determinants, using auxiliary fields (for details see
Ref.\cite{Emig:2003a}),
\begin{equation}
\label{logZ} \ln {\cal Z}_{\rm D} \,  =  \, - \, \frac{1}{2} \,
\ln \det \bbM_{\rm D},\quad \ln {\cal Z}_{\rm N} \,
=  \, - \, \frac{1}{2} \, \ln \det \bbM_{\rm N} \, \, .
\end{equation}
The kernels $\bbM_{\rm D}$ and $\bbM_{\rm N}$ are given
by
%\begin{subequations}
\label{gamma-kernel}
\begin{eqnarray}
[\bbM_{\rm D}]_{\alpha \beta}({\bf y},{\bf y}') \, & = & \,
[g_{\alpha}(y_1)]^{1/4} \, G[X_{\alpha}({\bf y})-X_{\beta}({\bf
y}')]
\, [g_{\beta}(y'_1)]^{1/4}\, , \label{AD} \\[2mm]
[\bbM_{\rm N}]_{\alpha \beta}({\bf y},{\bf y}') \, & = & \,
[g_{\alpha}(y_1)]^{1/4} \,
\partial_{n_\alpha(y_1)} \partial_{n_\beta(y'_1)}
G[X_{\alpha}({\bf y})-X_{\beta}({\bf y}')] \nonumber \\[0mm]
&&\times  [g_{\beta}(y'_1)]^{1/4} \label{AN} \, \, ,
\end{eqnarray}
%\end{subequations}
where $g_{\alpha}(y_1) = 1 + [h'_{\alpha}(y_1)]^2$
is the determinant of the induced metric, and
$n_\alpha(y_1)=(-1)^\alpha g_\alpha^{-1/2}(y_1)
[h'_\alpha(y_1),0,-1]$
is the normal vector to the surface $S_{\alpha}$, while
\begin{equation}
\label{G-free}
G({\bf x})=\frac{1}{4\pi^2}\frac{1}{{\bf x}^2}
\end{equation}
is the {\it free} Euclidean space Green's function with ${\bf x}=({\bf
  y},z)$. Eqs.~(\ref{cale}) to (\ref{G-free}) constitute the
functional determinant representation of the Casimir interaction. This
representation is exact. To proceed, the determinant has to be
evaluated either by perturbation theory in the deformation amplitude
or numerically for a specific shape of the surface profiles. First, we
present the perturbative approach.

For both boundary conditions $(X=D,N$) we devide by the partition
function ${\cal Z}_{{\rm X},\infty}$ for $H\to\infty$ and expand $\ln(
{\cal Z}_{\rm X}/{\cal Z}_{{\rm X},\infty})$ in a series $\ln(
{\cal Z}_{\rm X}/{\cal Z}_{{\rm X},\infty})|_0 + \ln(
{\cal Z}_{\rm X}/{\cal Z}_{{\rm X},\infty})|_1 + \ln(
{\cal Z}_{\rm X}/{\cal Z}_{{\rm X},\infty})|_2 + \ldots$, where the subscript indicates the
corresponding order in $h_\alpha$.  The lowest order result is
\begin{equation} \label{h0D}
\ln(
{\cal Z}_{\rm X}/{\cal Z}_{{\rm X},\infty})|_0 = \frac{AL}{H^3} \, \frac{\pi^2}{1440}
\end{equation}
for both types of modes, corresponding to two flat plates, cf.~Eq.
(\ref{eq:Casimir_E_plates}).  The first order result $\ln( {\cal
  Z}_{\rm X}/{\cal Z}_{{\rm X},\infty})|_1$ vanishes since we assume,
without loss of generality, that the mean deformations are zero, $\int
dy_1 h_{\alpha}(y_1) = 0$.  The second order
contribution is given by
\begin{eqnarray}
\label{zd} 
&& \ln({\cal Z}_{\rm X}/{\cal Z}_{{\rm X},\infty})|_2 \,  =  
\frac{\pi^2}{240} \, \frac{1}{H^5} \, \int d^3 y \,
\left\{ [h_1(y_1)]^2 + [h_2(y_1)]^2
\right\} \nonumber\\
& & \, - \, \frac{1}{2} \int d^3 y \int d^3 y' \,
K_{\rm X}({\bf y} - {\bf y}') \, \, \left\{  \frac{1}{2} \,
[h_1(y_1) - h_1(y'_1)]^2 +\frac{1}{2} \, [h_2(y_1) - h_2(y'_1)]^2
\right\} \nonumber\\
& & \, - \, \frac{1}{2} \int d^3 y \int d^3 y' \,
Q_{\rm X}({\bf y} - {\bf y}') \, \, [h_1(y_1) h_2(y'_1) +
h_2(y_1) h_1(y'_1)] \, \, . 
\end{eqnarray}
The terms in the first row are local contributions which are identical
for TM and TE modes. They follow also from a pairwise summation
approximation (PWS) that sums a ``renormalized'' Casimir-Polder potential
over the volumes of the interacting bodies \cite{Emig:2003a}. The
remaining terms are non-local and cannot be obtained in 
approximative schemes. For Dirichlet boundary conditions, the kernels
depend only on $|{\bf y}-{\bf y}'|$ and are given by 
%\begin{subequations}
\begin{eqnarray} \label{kdshort}
K_{\text{D}}(y) \, & = & \, -\frac{1}{2 \pi^4 y^8} + \frac{\pi^2}{128}
\, \frac{1}{H^6 y^2} \, \frac{\cosh^2(s)}{\sinh^6(s)},\\
Q_{\text{D}}(y) \, & = & \, \frac{\pi^2}{128} \, \frac{1}{H^6 y^2}
\, \frac{\sinh^2(s)}{\cosh^6(s)} \, \, ,
\end{eqnarray}
%\end{subequations}
where $s = \pi y / (2 H)$. The kernels for Neumann boundary conditions
assume a more complicated form since the normal derivative breaks the
equivalence of space and time directions. Hence, they depend
separately on $|y_0-y_0'|$ and $|{\bf y}_\| - {\bf y}'_\||$. Their
explicit form can be found in \cite{Emig:2003a}. The results obtained
so far apply to general uniaxial deformations of both surfaces.

\begin{figure}[h]
\includegraphics[width=9cm]{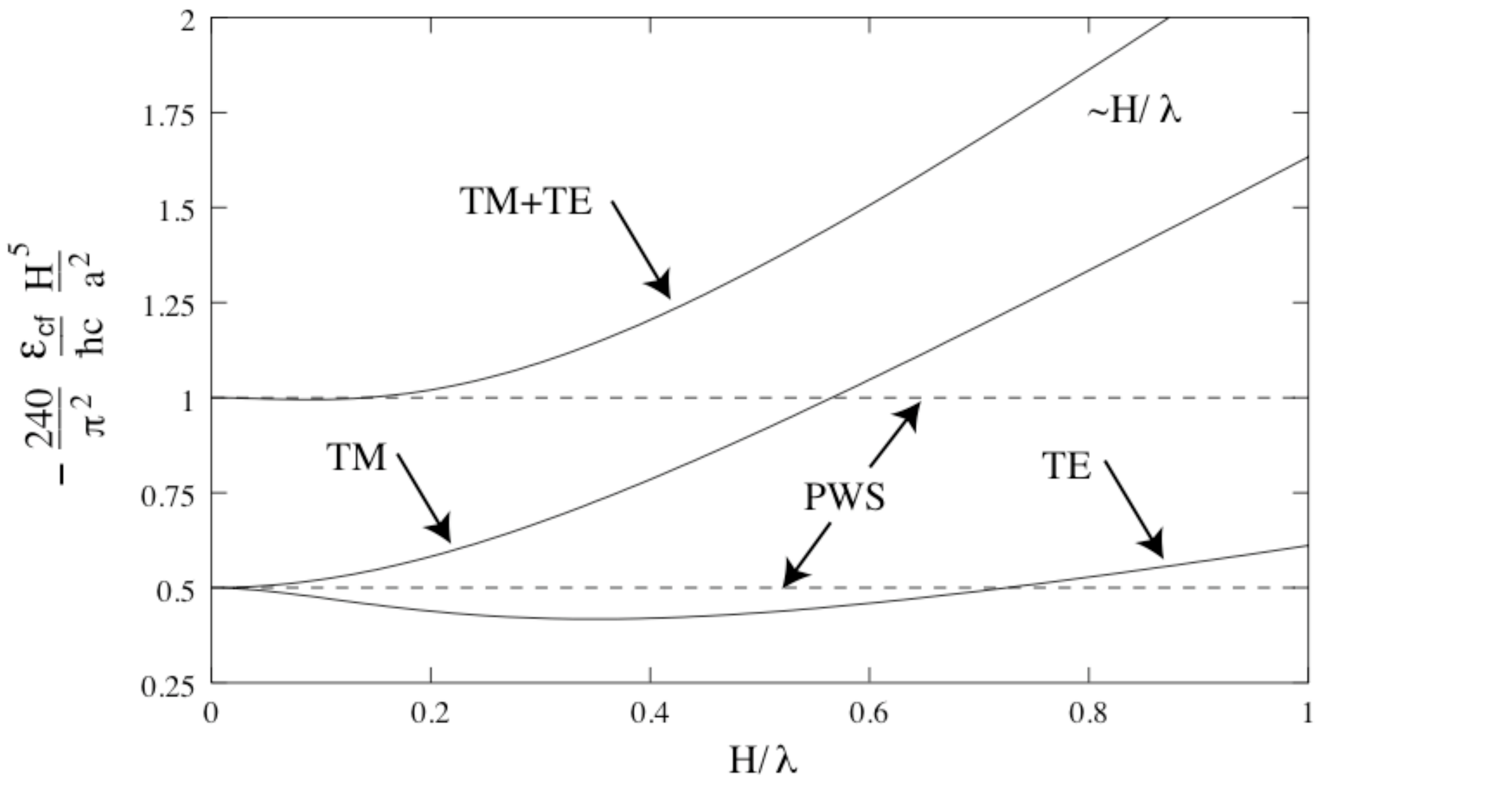}
\caption{\label{Fig:energy_corr-flat} Rescaled correction ${\cal
    E}_{\rm cf}$ to the Casimir energy due to the corrugation as given
  by Eq.~(\ref{ce+}) (upper curve).
  The lower curves show the separate contributions from TM and TE
  modes.  The rescaling of ${\cal E}_{\rm cf}$ is chosen such that the
  corresponding prediction of the pairwise summation (PWS)
  approximation [corresponding to the local terms of Eq.~(\ref{zd})]
  is a constant (dashed lines).}
\end{figure}

Now we apply these results to the important case of corrugated plates. 
We begin with the geometry depicted in Fig.~\ref{Fig:corr-flat} which
is parametrized by
\begin{equation} \label{corr}
h_1(y_1) \, = \, a \cos(2\pi y_1/\lambda) \, , \quad {\rm and}
\quad h_2(y_1) \, = 0 \, \, .
\end{equation}
For this profile, the computation of the partition function to second
order in $a$ reduces to the Fourier transforming of the kernels with
respect to $y_1$. The corresponding expression for ${\cal
E}$ in Eq.~(\ref{cale}) can be written as
\begin{equation}
\label{ce} {\cal E} \, = \, {\cal E}_0 + {\cal E}_{\rm cf} \, ,
\end{equation}
where $ {\cal E}_0$ is the energy per unit area of two flat plates
[see Eq.~(\ref{eq:Casimir_E_plates})] and
\begin{equation}
\label{ce+} 
 {\cal E}_{\rm cf} \, = \, - \,
\frac{\hbar c a^2}{H^5} \left[ G_{\rm
TM}\left(\frac{H}{\lambda}\right)+ G_{\rm
TE}\left(\frac{H}{\lambda}\right)\! \right] \, + \, {\cal O}(a^3) \, ,
\end{equation}
where the index cf of ${\cal E}_{\rm cf}$ stands for corrugated-flat
geometry. The functions that describe the $\lambda$-dependence in this
expression can be computed exactly \cite{Emig:2003a}. They can
be expressed in terms of the polylogarithm function 
${\rm Li}_n(z) =  \sum_{\nu=1}^\infty z^\nu/\nu^n$,
leading to
\begin{eqnarray}
&&G_{\rm TM}(x)=\frac{\pi^3x}{480}-\frac{\pi^2 x^4}{30} \ln(1-u) +
\frac{\pi}{1920 x} {\rm Li}_2(1-u) + \frac{\pi x^3}{24} {\rm
Li}_2(u)\nonumber\\
&+& \frac{x^2}{24} {\rm Li}_3(u) + \frac{x}{32\pi} {\rm Li}_4(u) + \frac{1}{64\pi^2} {\rm Li}_5(u)+\frac{1}{256\pi^3 x} \left(
{\rm Li}_6(u)-
\frac{\pi^6}{945}\right)\, , \label{tm} \nonumber\\
\\
&&G_{\rm TE}(x)=\frac{\pi^3 x}{1440}-\frac{\pi^2 x^4}{30} \ln(1-u)
+\frac{\pi}{1920 x} {\rm Li}_2(1-u) \nonumber\\
&-&\frac{\pi
x}{48}\left(1+2x^2\right){\rm Li}_2(u)
+\left(\frac{x^2}{48}-
\frac{1}{64}\right){\rm Li}_3(u)+
+\frac{5x}{64\pi}{\rm Li}_4(u)\nonumber\\
&+&\frac{7}{128\pi^2} {\rm
Li}_5(u) +\frac{1}{256\pi^3x}\left(\frac{7}{2}{\rm Li}_6(u)-\pi^2
{\rm Li}_4(u) +\frac{\pi^6}{135}\right) \label{te} 
\end{eqnarray}
with $u \equiv \exp(-4\pi x)$.
Figure~\ref{Fig:energy_corr-flat} displays separately the
contributions from $G_{\rm TM}$ and $G_{\rm TE}$ to the corrugation
induced correction ${\cal E}_{\rm cf}$ to the Casimir energy.
While $G_{\rm TM}(H/\lambda)$ is a monotonically increasing function
of $H/\lambda$, $G_{\rm TE}(H/\lambda)$ displays a minimum for
$H/\lambda \approx 0.3$. 

Examining the limiting behaviors of Eq.~(\ref{ce+}) is instructive.
In the limit $\lambda \gg H$, the functions $G_{\rm TM}$ and
$G_{\rm TE}$ approach constant values, and the total Casimir energy
takes the $\lambda$-independent form
\begin{equation}
{\cal E}=-\frac{\hbar c}{H^3}\frac{\pi^2}{720}
\left(1+3\frac{a^2}{H^2} \right)
\, \, + \, {\cal O}(a^3) \, .
\end{equation}
Note that {\em only} in this case both wave types provide the same
contribution to the total energy and the result agrees with the
pairwise summation approximation (see
Fig.~\ref{Fig:energy_corr-flat}).  In the opposite limit of $\lambda
\ll H$ both $G_{\rm TM}$ and $G_{\rm TE}$ grow
linearly in $H/\lambda$.  Therefore, in this limit the correction to
the Casimir energy decays according to a {\em slower} power law in
$H$, as
\begin{equation}
\label{large-H} {\cal E}=-\frac{\hbar c}{H^3}\frac{\pi^2}{720}
\left(1+2\pi\frac{a^2}{\lambda H}\right)
\, \, + \, {\cal O}(a^3) \, ,
\end{equation}
with an amplitude proportional to $1/\lambda$.  Note that this
behavior is completely missed by the pairwise summation approach which
always yields a $\lambda$ independent Casimir energy in the presence
of modulations on one plate \cite{Emig:2003a}. As we will discuss
below in the context of the numerical approach, the apparent
divergence for $\lambda\to 0$ in Eq.~(\ref{large-H}) is an artifact of
the perturbative expansion which assumes that the amplitude $a$ is
the smallest length scale.

Next we turn to a numerical approach for computing the functional
determinants of Eq.~(\ref{logZ}). Such an approach has been developed
for periodic surface profiles in \cite{Emig:2003b,Buscher:2004a}.  In
this approach, it is convenient to compute directly the Casimir force
$F=-\partial_H {\cal E}$ per unit area which is the sum of TM and TE
contributions, $F=F_\text{TM}+F_\text{TE}$ that according to
Eq.~(\ref{logZ}) are given by (for $X=D, N$)
\begin{eqnarray}
\label{eq:force}
F_{\rm X}=-\frac{\hbar c}{2AL}\,{\rm Tr}\left(\bbM_{\rm X}^{-1}
\partial_H\bbM_{\rm X}\right) \, .
\end{eqnarray}
The right hand side of this expression is always finite, and no
divergences due to self-energies have to be subtracted. The trace in
Eq.~(\ref{eq:force}) can be computed efficiently by Fourier
transforming $\bbM$ with respect to ${\bf y}$, ${\bf y}'$. The so
transformed operator can then be transformed to block-diagonal form by
making use of the periodicity of the surface profile along the $y_1$
direction. In this representation the blocks can be numbered by the
wave vector $q_1 \in [0,2\pi/\lambda)$ along the $y_1$ direction.  A
block matrix with label $q_1$ couples only waves whose momenta differ
from the Bloch momentum $q_1$ by integer multiples of $2\pi/\lambda$.
The integers multiplying $2\pi/\lambda$ number the matrix elements
within a block matrix. Hence, the problem of computing the total trace has been
simplified to the computation of the trace of each block matrix with
label $q_1$. Finally, integration over $q_1$ from $0$ to
$2\pi/\lambda$ and over the unrestricted momenta $q_0$, $q_2$ (along
the time direction and invariant spatial direction of the surfaces,
respectively) yields the force of Eq.~(\ref{eq:force}). For the
particular choice of a {\it rectangular} corrugation
[see Fig.~\ref{Fig:lateral_vs_delta}(a)] analytic expressions for all
matrix elements of $\bbM$ can be obtained. For details of the
implementation of the numerical approach and expressions for the
matrix elements see \cite{Buscher:2004a}. 

\begin{figure}[h]
\includegraphics[width=7cm]{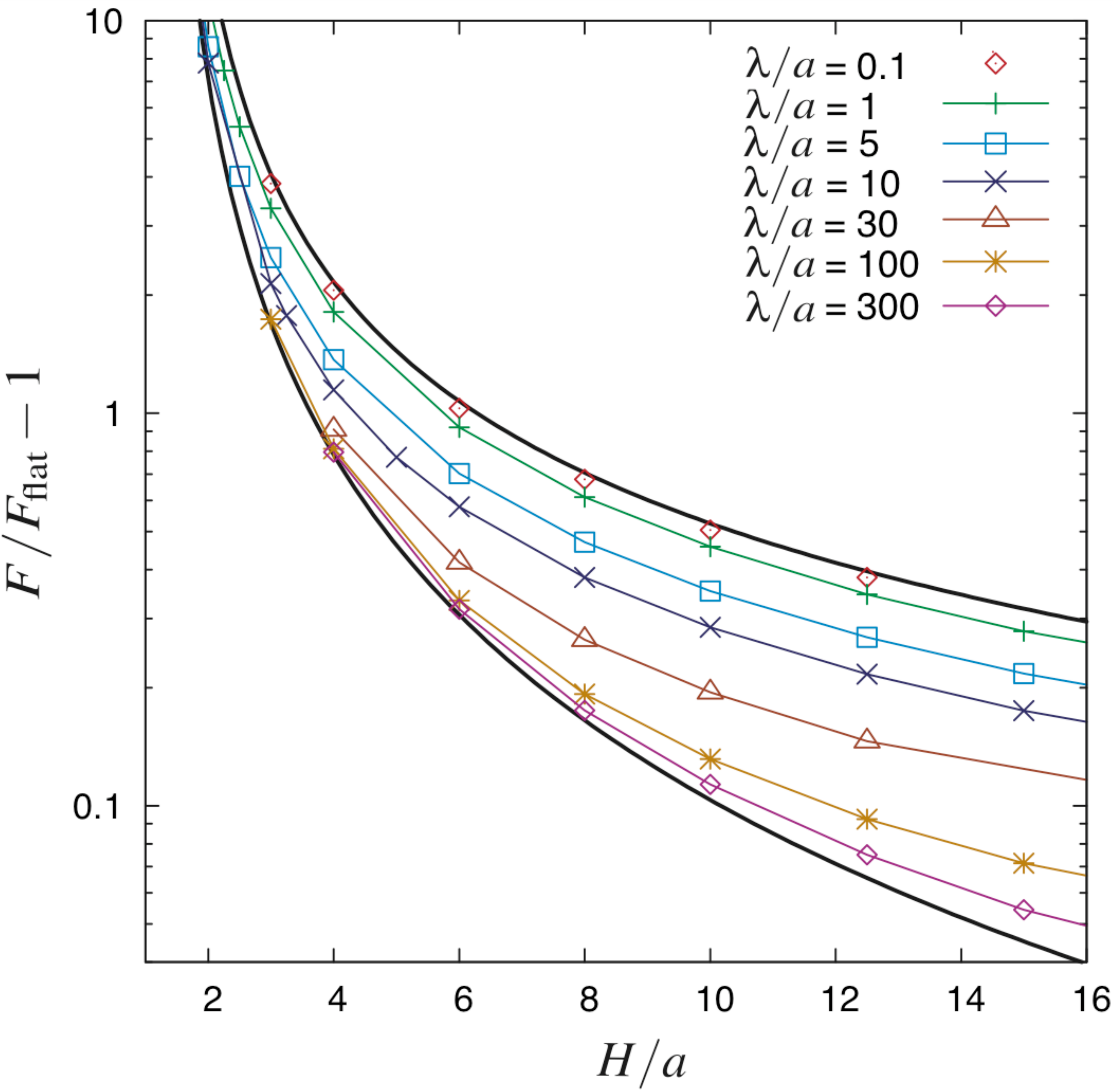}
\caption{\label{Fig:rect_corr_normal} Total Casimir force as function
  of the mean plate separation $H$. Shown is the relative change of
  the force compared to the total Casimir force $F_{\rm flat}$ between
  two flat plates. The two bold curves enclosing the numerical data
  are the analytical results $F_0$ for $\lambda \to 0$ (upper curve)
  and $F_\infty$ for $\lambda \to \infty$ (lower curve), see text.}
\end{figure}

In analogy to the profile of Fig.~\ref{Fig:corr-flat} we consider the
corresponding situation of a flat plate and a plate with a rectangular
corrugation profile parametrized by
\begin{equation}
\label{eq:profile}
h_1(y_1)\:=\:\left\{
\begin{array}{lll}
+a & \text{for} & |y_1| < \lambda/4\\
-a & \text{for} & \lambda/4 < |y_1| < \lambda/2
\end{array}\right.,
\end{equation}
and continuation by periodicity $h_1(y_1)=h_1(y_1+n\lambda)$ for any
integer $n$. The numerical results for
the total Casimir force between the two plates is shown in
Fig.~\ref{Fig:rect_corr_normal} for different corrugation wavelengths
$\lambda$. For all $\lambda$, the forces at a fixed separation $H$ are
bounded between a minimal force $F_\infty$ and a maximal force $F_0$.
For small $\lambda/a$ the upper bound $F_0$ is approached whereas for
asymptotically large $\lambda/a$ the force converges towards the lower
bound $F_\infty$. Analytic expressions can be derived for these bounds.
For large $\lambda$, the corrugated surface is composed of large flat
segments with a low density of edges. At sufficiently small surface
separations $H \ll \lambda$ the main contribution to the force comes
from wavelengths which are much smaller than the scale $\lambda$ of
the surface structure. Thus in the dominant range of modes diffraction
can be neglected, and a simple proximity force approximation
\cite{Parsegian:2005eu} should be applicable.  Such an approximation
assumes that the total force can be calculated as the sum of local
forces between opposite {\it flat} and {\it parallel} small surface
elements at their local distance $H-h_1(y_1)$. No distinction is made
between TM and TE modes.  This procedure is rather simple for the
rectangular corrugation considered here since the surface has no
curvature (except for edges). There are only two different distances
$H+a$, $H-a$ which contribute one half each across the entire surface
area, leading for $\lambda \to \infty$ to the proximity approximation
for the force,
\begin{equation}
\label{eq:F_infty}
F_\infty/A\:=\:-\frac{\pi^2\hbar c}{240}\,
\frac{1}{2}
\left[\frac{1}{(H-a)^4}+\frac{1}{(H+a)^4}\right]\, .
\end{equation}
In the limit $\lambda\to 0$ the important fluctuations should not get
into the narrow valleys of the corrugated plate. Even for small but
finite $\lambda$ this picture should be a good, though approximate,
description since it still effects the wavelengths of order $H$ which
give the main contribution to the force. Thus one can expect that the
plates feel a force which is equal to the force between two {\it flat}
plates at the {\it reduced} distance $H-a$.  Fortunately, this
expectation can be checked by an explicit calculation since the
leading part of determinant of $\bbM_X$ in the limit $\lambda\to 0$
can be computed. Indeed, this computation confirms the expectation,
leading to the Casimir force per surface area \cite{Buscher:2004a}
\begin{equation}
\label{eq:F_0}
F_0/A \:=\: -\frac{\pi^2}{240}\frac{1}{(H-|a|)^4}
\end{equation}
with equal contributions from TM and TE modes.  Notice that this
result is not analytic in $a/H$ and is {\it exact} in the limit
$\lambda \to 0$.  As we have seen before, perturbation theory for
smoothly deformed surfaces always yields corrections to the
interaction of order $a^2$. However, for small $a/H$, the result of
Eq.~(\ref{eq:F_0}) has the expansion
\begin{equation}
F_0/A \:=\: - \frac{\pi^2}{240}\frac{1}{H^4}
\left[1+4\frac{|a|}{H}+{\cal O}\left(\left(\frac{a}{H}\right)^2\right)\right]
\end{equation}
which indicates that perturbation theory is not applicable if $\lambda
\ll a$. This implies that the apparent divergent behavior for $\lambda\to 0$ in
Eq.~(\ref{large-H}) actually disappears for $\lambda\simeq a$. 

\subsection{Lateral forces}
\label{sec:lateral_force}

\begin{figure}[h]
\includegraphics[width=7cm]{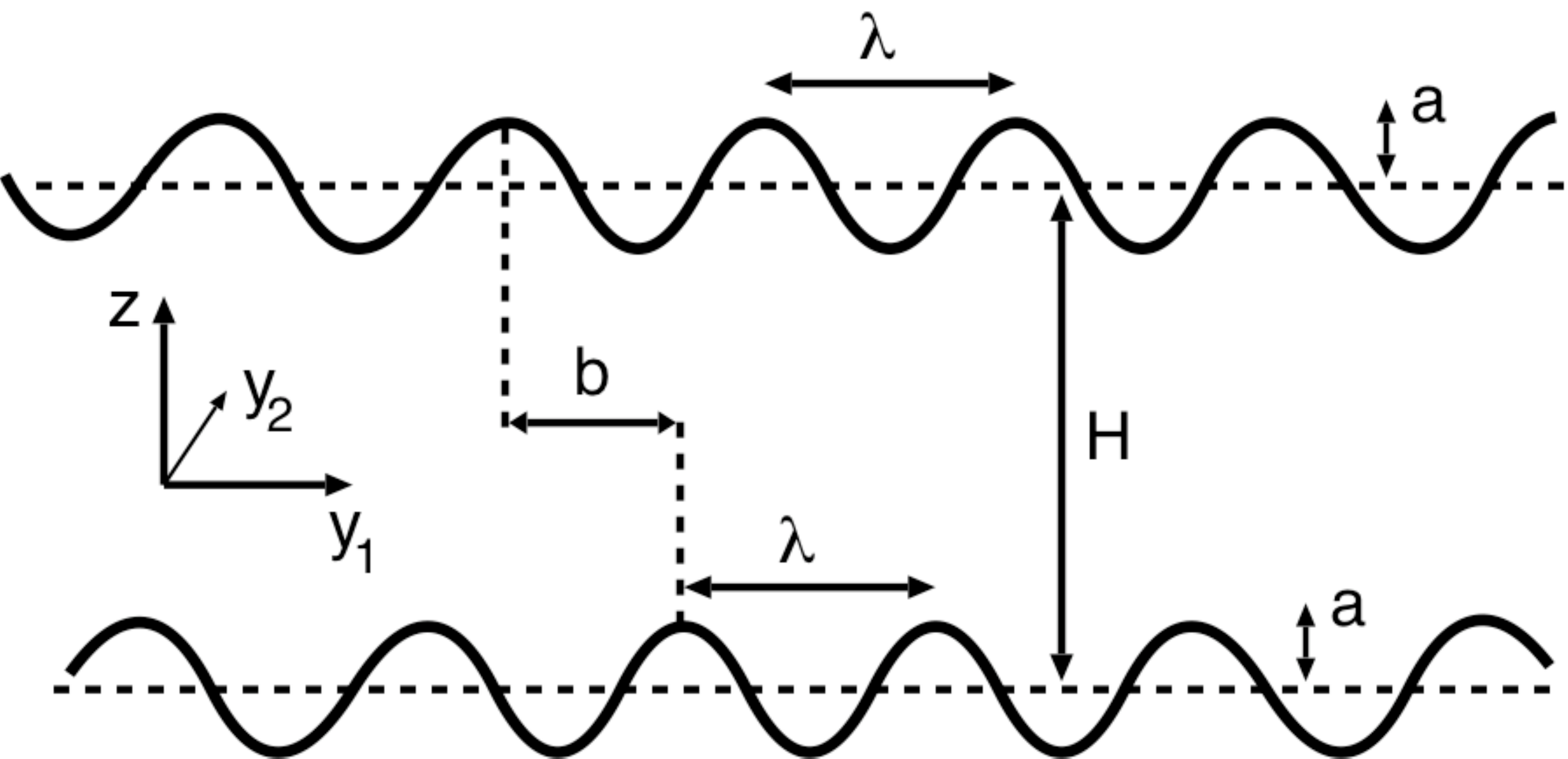}
\caption{\label{Fig:corr-corr}Geometry used for calculating the
  lateral Casimir force between two corrugated plates with lateral
  shift $b$. The equilibrium position is at $b=\lambda/2$. }
\end{figure}

As a natural generalization of the geometry of the previous section,
we study the Casimir interaction between two sinusoidally corrugated
plates. For direct correspondence to experiments for this type of
configuration \cite{Chen:2002d}, we consider the specific profiles
\begin{equation}
\label{corr-2} h_1(y_1) \, = \, a \cos(2\pi y_1/\lambda) \, ,
\quad {\rm and} \quad h_2(y_1) \, =  \, a \cos\left(2\pi
(y_1+b)/\lambda\right) \, \, ,
\end{equation}
which are shifted relative to each other by the length $b$ (see
Fig.~\ref{Fig:corr-corr}). When these profiles are substituted into
the general expression for the second order term of the partition
function of Eq.~(\ref{zd}) one finds for the Casimir energy
\begin{equation}
{\cal E} \, = \, {\cal E}_0 + 2 {\cal E}_{\rm cf} + {\cal E}_{\rm
cc},
\end{equation}
with ${\cal E}_{\rm cf}$ given in Eq.~(\ref{ce+}), and where the
corrugation-corrugation interaction energy ${\cal E}_{\rm cc}$ can be
calculated in terms of the kernels $Q_{\rm X}({\bf y})$ in
Eq.~(\ref{zd}). Besides oscillating contributions to the normal
Casimir force from ${\cal E}_{\rm cc}(b)$, a {\em lateral} force
\begin{equation}
\label{f-lat} F_{\rm lat}= - \frac{\partial {\cal E}_{\rm
cc}}{\partial b}\, ,
\end{equation}
is induced by the corrugation-corrugation interaction.  This lateral
force is much better suited for experimental tests of the influence of
deformations, since there is no need for subtracting a larger baseline
force (the contribution of flat plates) as in the case of the normal
force. The lateral force can be also employed as a actuation mechanism
in mechanical oscillators as we will see in Section
\ref{sec:driven_system}. In analogy to the previous section, the 
corrugation-corrugation interaction can be expressed as
\begin{equation}
\label{Ecc} {\cal E}_{\rm cc} \, = \, \frac{\hbar c a^2}{H^5}
\cos\left(\frac{2\pi
 b}{\lambda}\right)
 \left[ J_{\rm TM}\left(\frac{H}{\lambda}\right)+
J_{\rm TE}\left(\frac{H}{\lambda}\right)\! \right] \, + \, {\cal O}(a^3) 
\end{equation}
with 
%\begin{subequations}
\begin{eqnarray}
\label{j-tm} 
&&J_{\rm TM}(x) = 
\frac{\pi^2}{120}\left(16x^4-1\right){\rm arctanh} (\sqrt{u})+
\sqrt{u}\left[\frac{\pi}{12}\left(x^3-\frac{1}{80x}\right)\Phi(u,2,\frac{1}{2})
\right. \nonumber\\
&& \left.
+\frac{x^2}{12}\, \Phi(u,3,\frac{1}{2}) + \frac{x}{16\pi}\, \Phi(u,4,\frac{1}{2})
+\frac{1}{32\pi^2} \, \Phi(u,5,\frac{1}{2}) +\frac{1}{128\pi^3 x}
\, \Phi(u,6,\frac{1}{2})\right]\, , \nonumber\\
\\
\label{j-te} 
&& J_{\rm TE}(x)  = 
\frac{\pi^2}{120}\left(16x^4-1\right){\rm arctanh} (\sqrt{u})+
\sqrt{u}\left[-\frac{\pi}{12}\left(x^3+\frac{x}{2}+\frac{1}{80x}\right)
\right. \nonumber\\
&& \left. \times
\Phi(u,2,\frac{1}{2})
+\frac{1}{24}\left(x^2-\frac{3}{4}\right)\Phi(u,3,\frac{1}{2})
 + \frac{5}{32\pi} \left( x - \frac{1}{20x}\right)
\Phi(u,4,\frac{1}{2})
\right. \nonumber\\
&& \left.
 + \frac{7}{64\pi^2} \, \Phi(u,5,\frac{1}{2})
+\frac{7}{256\pi^3 x} \, \Phi(u,6,\frac{1}{2}) \right] \, ,
\end{eqnarray}
%\end{subequations}
where $u\equiv \exp(-4\pi x)$ and $\Phi(z,s,a)=\sum_{k=0}^\infty
z^k/(a+k)^s$ is the Lerch transcendent. In the limit of large
corrugation length, $H/\lambda\to 0$, this result agrees to lowest
order with a pairwise summation approximation where $J_{\rm
  TM}(0)+J_{\rm TE}(0)=\pi^2/120$. 

At the other extreme of $\lambda \ll H$, $J_{\rm TM}(x)+J_{\rm TE}(x)$
decays {\it exponentially} fast. This decay distinguishes the lateral
force from the normal force.  In particular, for large $x=H/\lambda$,
we get to leading order
\begin{equation}
J_{\rm TM}(x)+J_{\rm TE}(x) = \frac{4\pi^2}{15}\left( x^4 + {\cal
O}(x^2)\right) e^{-2\pi x} \,  .
\end{equation}
Since $J_{\rm TM}(x)+J_{\rm TE}(x)$ is positive for all values of $x$,
the equilibrium position of two modulated surfaces is predicted at
$b=\lambda/2$.  This corresponds to aligning the maxima and minima of
the two corrugations (cf.~Fig.~\ref{Fig:corr-corr}).

\begin{figure}[h]
\includegraphics[width=10cm]{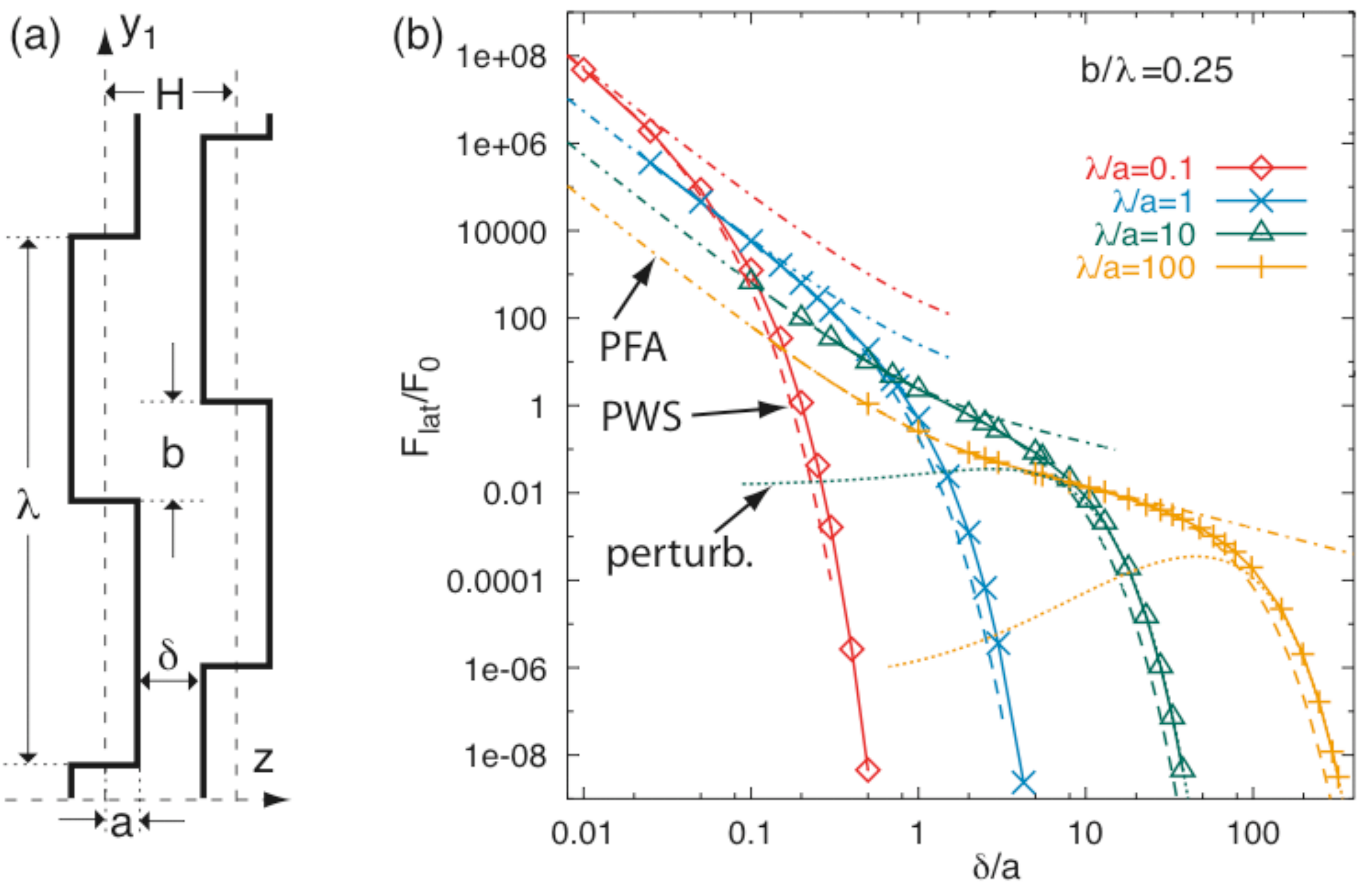}
\caption{\label{Fig:lateral_vs_delta} (a) Geometry consisting of two
  parallel plates with laterally shifted uniaxial rectangular
  corrugations. (b) Lateral force $F_{\rm lat}$ (in units of normal
  force $F_0$ between flat surfaces) at $b=\lambda/4$ for the geometry
  shown in (a) as function of the gap $\delta$ (solid curves).
  Plotted are also the proximity force (PFA, dash-dotted curves) and
  pairwise summation (PWS, dashed curves) approximations, and the
  perturbative result $F_{\rm pt}$ that follows from a calculation for
  sinusoidal profiles (dotted curves).}
\end{figure}
\begin{figure}[h]
\includegraphics[width=7cm]{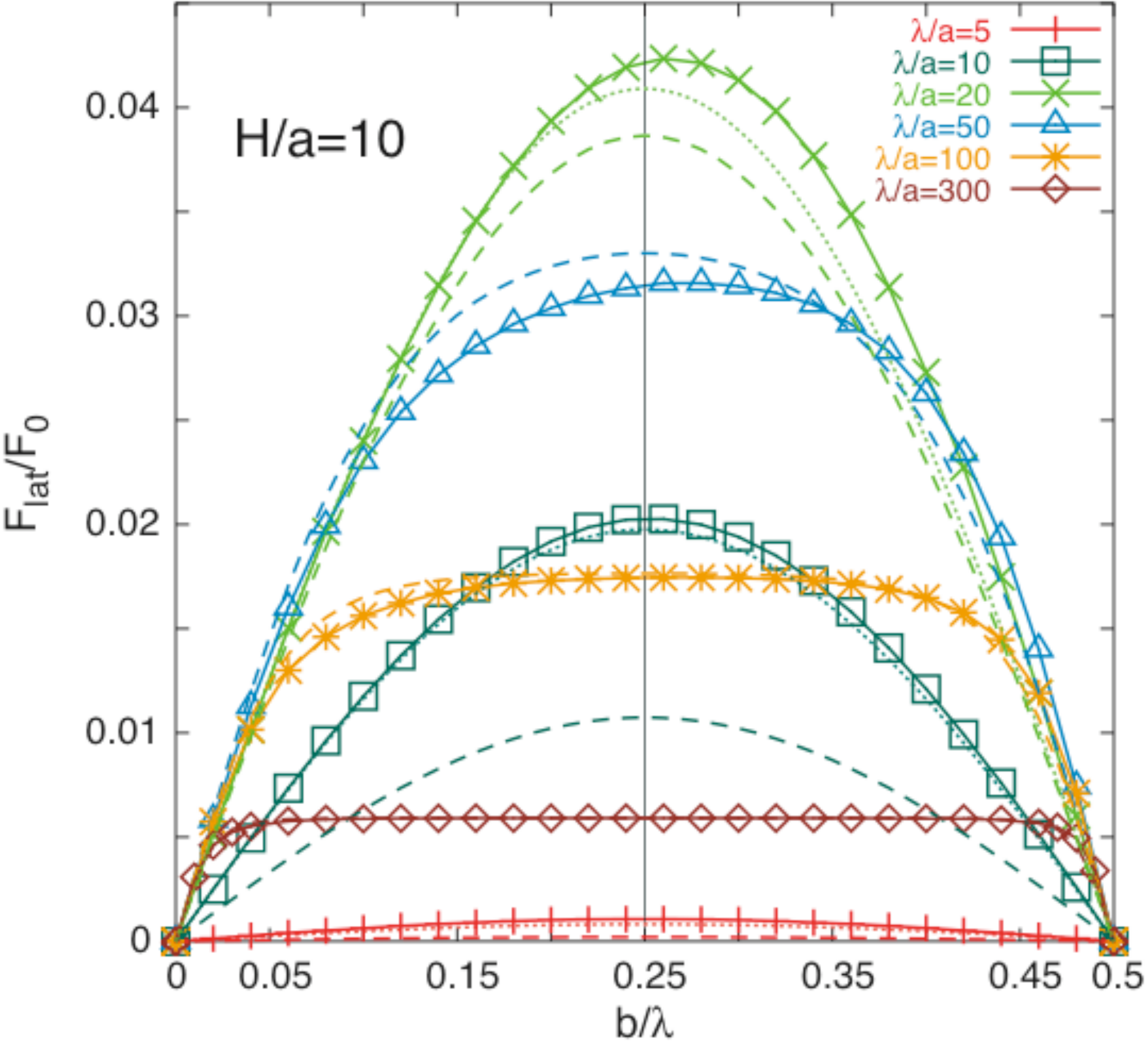}
\caption{\label{Fig:lateral_vs_b}Shape dependence of $F_{\rm lat}$ on
  the lateral surface shift $b$ at fixed distance $H=10a$ for
  different corrugation lengths. The dashed and the dotted curves
  represent the PWS and the full perturbative result for sinusoidal
  profiles with arbitrary $H/\lambda$, respectively.  }
\end{figure}

The numerical approach for computing the functional determinant in the
case of periodic surfaces can be also applied to the lateral force
\cite{Buscher:2005a}.  We consider again a rectangular corrugation but
now on both surfaces with a lateral shift of $b$,
cf.~Fig.~\ref{Fig:lateral_vs_delta}. The numerical results for the
lateral force in this geometry are summarized in
Figs.~\ref{Fig:lateral_vs_delta} and
\ref{Fig:lateral_vs_b}. Fig.~\ref{Fig:lateral_vs_delta} shows the
numerical result for the lateral force for a shift $b=\lambda/4$ and
different values of $\lambda/a$ over more than 4 orders of magnitude
for the gap $\delta=H-2a$, together with two approximate results (PFA
and PWS) and the perturbative result for sinusoidal profiles for
$\lambda\ll H$. An exponential decay of the force as predicted by
perturbation theory can be clearly observed.

The PFA yields a lateral force per unit area $F_{\rm lat,PFA}=[2{\cal
  E}_0(H)-{\cal E}_0(H-2a)-{\cal E}_0(H+2a)]/\lambda$ for
$0<b<\lambda/2$ where ${\cal E}_0$ has the same meaning as before.
$F_{\rm lat,PFA}$ changes sign at $b=\lambda/2$ discontinuously which
is an artifact of this approximation. The pairwise summation (PWS) of
Casimir-Polder potentials is strictly justified for rarefied media
only but it is often also applied to metals, using the two-body
potential $U(r)=-(\pi/24)\hbar c/r^7$ with the amplitude chosen such
as to reproduce the correct result for flat ideal metal plates
\cite{Bordag:2001b}. It yields a lateral force $F_{\rm
  lat,PWS}=-\frac{\partial}{\partial b} \int_{V_l} d^3 {\bf x}
\int_{V_r} d^3 {\bf x}' U(|{\bf x}-{\bf x}'|)$ with $V_l$ and $V_r$
denoting the semi-infinite regions to the left and right of the two
surfaces in Fig.~\ref{Fig:lateral_vs_delta}(a), respectively.  $F_{\rm
  lat,PWS}$ can be obtained by numerical integration.  For small gaps
$\delta$, both approximations agree and match the exact numerical
results.  Beyond $\delta \gtrsim \lambda/20$ the PFA starts to fail
since it does not capture the exponential decay of $F_{\rm lat}$ for
increasing $\delta$. The PWS approach has a slightly larger validity
range and reproduces the exponential decay. However it deviates by at
least {\it one order of magnitude} from $F_{\rm lat}$ for $\delta
\gtrsim 2.5 \lambda$.

Although the perturbative result of Eq.~(\ref{Ecc}) applies to
sinusoidal surfaces, it is instructive to compare it to the numerical
results for the rectangular profiles. Since the lateral force decays
exponentially, $F_{\rm lat} \sim e^{-2\pi H/\lambda}$, with the
characteristic scale set by the modulation wavelength of the profile,
the force at large $H$ should be determined by the lowest harmonic of
the periodic surface profile. This implies a {\it universal} lateral
force for large $H\gg \lambda$ that is independent of the precise form
of the surface corrugation. This universal force is the force between
two sinusoidal surfaces where the amplitude follows from the
projection of an arbitrary periodic profile of wavelength $\lambda$
onto a sinusoidal profile with the same wavelength. The latter force
follows from Eq.~(\ref{Ecc}) and in the limit $\lambda\ll H$ is given
by
\begin{equation}
\label{eq:f-pert}
F_{\rm pt}=\frac{8\pi^3\, \hbar c}{15} \frac{a_0^2 A}{\lambda^5 H}\,
\sin\left(\frac{2\pi}{\lambda} \, b\right) e^{-2\pi H/\lambda} \, ,
\end{equation}
where we assumed an amplitude $a_0$ for the sinusoidal profiles.  For
the rectangular corrugation of Fig.~\ref{Fig:lateral_vs_delta} the
lowest harmonic has the amplitude $a_0=4a/\pi$. When we compare
$F_{\rm pt}$ and the numerical results of
Fig.~\ref{Fig:lateral_vs_delta}(b) we find indeed excellent agreement
for distances $\delta \gtrsim \lambda$.

The universal behavior of the lateral force is also clearly
demonstrated by the dependence of the lateral force on the surface
shift $b$. Corresponding numerical results together with PWS
approximations and the force that follows from the {\it full}
perturbative result of Eq.~(\ref{Ecc}) for sinusoidal surfaces with
arbitrary $H/\lambda$ are shown in Fig.~\ref{Fig:lateral_vs_b} for
fixed $H=10a$ and varying $\lambda/a$.  With decreasing $\lambda$,
three regimes can be identified. For $\lambda \gg H$, the force
profile resembles almost the rectangular shape of the surfaces, and
the PWS approximation yields consistent results. For smaller
$\lambda$, yet larger than $H$, the force profile becomes asymmetric
with respect to $b=\lambda/4$ and more peaked, signaling the crossover
to the universal regime for $\lambda \lesssim H$ where the force
profile becomes sinusoidal. In the latter case, for not too small
$\lambda/a \approx 10$, the numerical results for $F_{\rm lat}$ indeed
agree well with the perturbative result for sinusoidal surfaces with
arbitrary $H/\lambda$.  We note that the PWS approach fails to predict
the asymmetry of the force profile, and the PFA even predicts no
variation with $b$ for $0<b<\lambda/2$. To observe this universal
behavior of the lateral force experimentally, one should consider
surfaces with very small corrugation wavelengths in the range of
nanometers so that the exponential decay does not diminish the force
for $H\gg \lambda$ too strongly.

\subsection{Cylinders}

In this Section we give examples for two central aspects of
fluctuation forces: Effects resulting from the non-additivity and the
particular properties of systems with a codimension of two which plays
a special role as we will see below.  These problems are considered in
the context of interactions between cylinders and sidewalls. It has
been demonstrated that Casimir forces in these geometries have only a
weak logarithmic dependence on the cylinder radius \cite{Emig:2006a}
and can be non-monotonic
\cite{Rodriguez:2007a,Rahi:2008bv,Rahi:2008kb} -- consequences of
codimension and non-addivity. These forces between
quasi-one-dimensional structures could be probed in mechanical
oscillators that are composed of nano-wires or carbon nanotubes.
Exact results for the interaction can be obtained by employing a
recently developed scattering approach for Casimir forces
\cite{Emig:2007os,Emig:2008ee}. This approach is based on the concept
that electromagnetic  Casimir interactions result from fluctuating 
currents inside the bodies. It is possible to formulate an effective action
for the multipole moments ${\bf Q}_{\alpha,X}$ of the currents inside the bodies
where $\alpha$ labels the bodies and $X$ is a multi-index that numbers
polarizations (electric and magnetic multipoles) and  the elements of the
basis for the multipole expansion, e.g., cylindrical waves. The effective
action can then be written as the quadratic form
\begin{equation}
  \label{eq:EM-total_action-final}
  S = \sum_{\alpha,\alpha'} \sum_{X,X'} {\bf Q}^*_{\alpha,X} {\mathbb M}_{\alpha\alpha', XX'} {\bf Q}_{\alpha',X'} \, ,
\end{equation}
with the matrix kernel
\begin{equation}
  \label{eq:EM-def-matrix-M}
  {\mathbb M}_{\alpha\alpha',XX'}= \kappa \left\{ \left[({\mathbb T}_\alpha )^{-1} \right]_{XX'}
\delta_{\alpha\alpha'} - {\mathbb U}_{\alpha\alpha',XX'} (1-\delta_{\alpha\alpha'})
\right\} \, ,
\end{equation}
where $\kappa$ is the Wick rotated frequency, $\omega=ic\kappa$, the
matrix ${\mathbb T}_\alpha$ is the so-called T-matrix of object
$\alpha$ that relates incoming and scattered waves and ${\mathbb
  U}_{\alpha\alpha'}$ is a ``translation'' matrix that relates the
incoming wave at object $\alpha$ to the outgoing wave at object
$\alpha'$. The T-matrix is related to the scattering matrix of the
object, ${\mathbb S}_\alpha$, by the relation ${\mathbb
  T}_\alpha=({\mathbb S}_\alpha-1)/2$. Analytic results for all
elements of the scattering matrix are available for symmetric shapes
such as cylinders and spheres. The ${\mathbb S}_\alpha$ matrix
contains all information about shape and material composition of the
object that is relevant to the Casimir interaction. The translation
matrices ${\mathbb U}_{\alpha\alpha'}$ are independent of the
properties of the interacting bodies and depend only on the relative
position (separation vector) of the objects $\alpha$ and $\alpha'$ and
the properties of the fluctuating field. For the electromagnetic
field, the translation matrices are known in many bases, e.g., for
cylindrical and spherical waves \cite{Emig:2008ee}. To obtain the
Casimir energy, the multipole fluctuations are integrated out, leading
to the determinant of the infinite dimensional matrix ${\mathbb
  M}$. Integration over all frequencies $\kappa$ yields the
interaction energy
\begin{equation}
  \label{eq:EM-Casimir-energy-gen-result}
  {\cal E} = \frac{\hbar c}{2\pi} \int_0^\infty d\kappa \ln\frac{\det {\mathbb M}}
{\det {\mathbb M}_\infty}  \, ,
\end{equation}
where the division by the determinant of the matrix ${\mathbb
  M}_\infty$ accounts for the subtraction of the residual energy of
the configuration where the separations between all objects tend to
infinity. Since the translation matrices decay to zero with increasing
separation, the matrix ${\mathbb M}_\infty$ is given by
Eq.~(\ref{eq:EM-def-matrix-M}) with the ${\mathbb U}_{\alpha\alpha'}$
set to zero. In the special case of two objects, the energy can be
simplified to \cite{Emig:2007os}
\begin{equation}
  \label{eq:energy_gen}
  {\cal E}_2 = \frac{\hbar c}{2\pi} \int_0^\infty d \kappa 
\ln \det (1 - \mathbb{N} ) \, ,
\end{equation}
where $\mathbb{N}=\mathbb{T}_1\mathbb{U}_{12}\mathbb{T}_2\mathbb{U}_{21}$.

\begin{figure}[h]
\includegraphics[width=8cm]{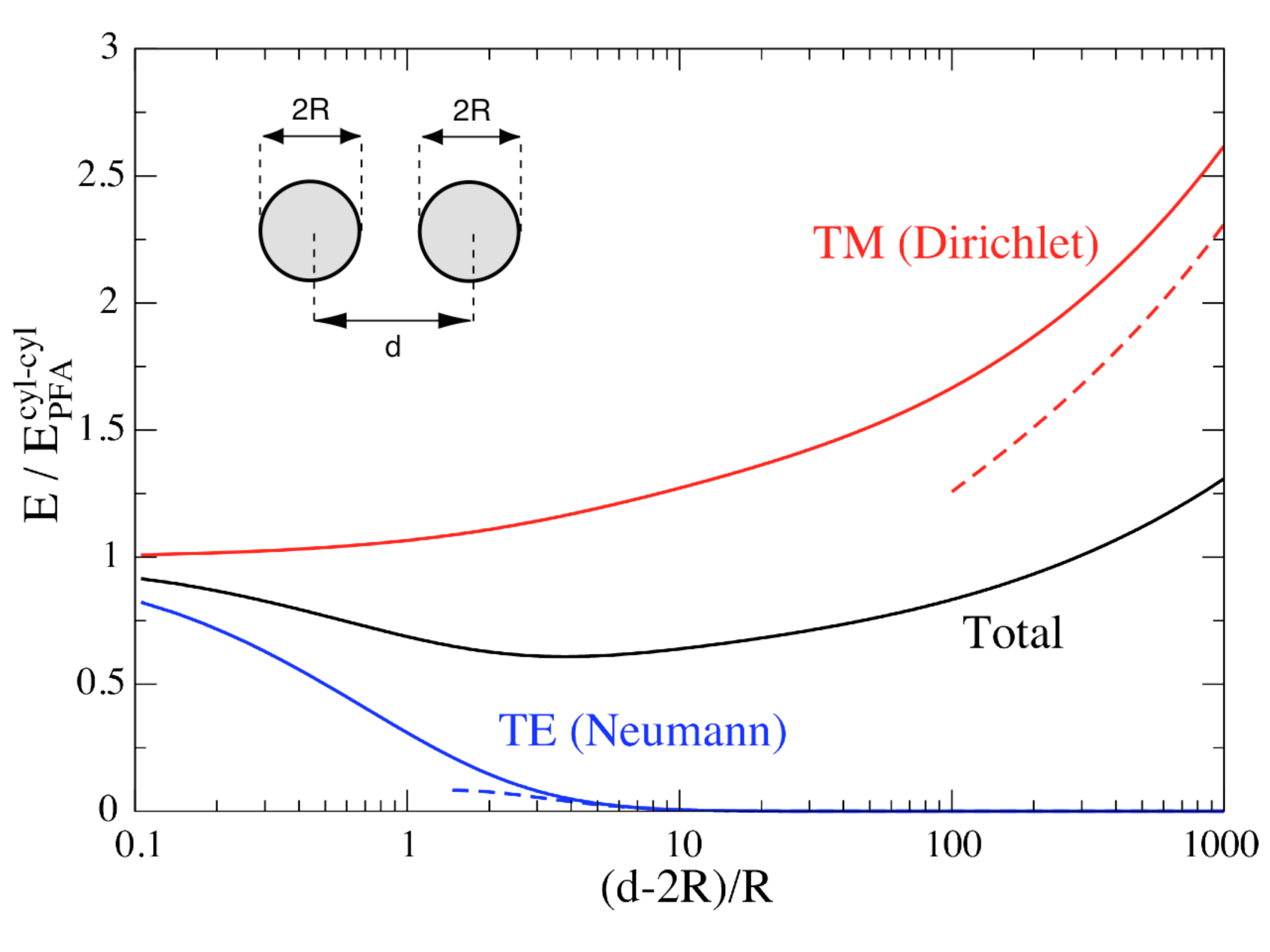}
\caption{\label{Fig:2cylinders}Casimir energy for two cylinders of
  equal radius $R$ as a function of surface-to-surface distance $d-2R$
  (normalized by the radius). The energy is divided by the PFA
  estimate $E_{\rm PFA}^{\rm cyl-cyl}=-\frac{\pi^3}{1920} \hbar c
  L\sqrt{R/(d-2R)^5}$ for the energy which is applicable in the limit
  $d\to 2R$ only. The solid curves show numerical results; the dashed lines
  represent the asymptotic results of
  Eq.~(\ref{eq:E-two-cylinders}). The inverse logarithmic correction to the
  leading order result for TM modes cause very slow convergence. For
  the parameter range shown here it was sufficient to consider $m=40$
  partial waves to obtain convergence.}
\end{figure}

First, the scattering approach is applied to two parallel, infinitely
long, perfectly conducting cylinders of equal radius $R$ and
center-to-center separation $d$, see Fig.~\ref{Fig:2cylinders}.  For
this geometry it is most convenient to use cylindrical vector waves
for the multipole expansion. This basis consists of the vector fields
${\bf M}^{i(o)}_{k_z m}({\bf x}) = \frac{1}{q} \nabla\times {\bf
  V}^{i(o)}({\bf x})$ for magnetic (M) multipoles and ${\bf
  N}^{i(o)}_{k_z m}({\bf x}) = \frac{c}{q\omega}
\nabla\times\nabla\times {\bf V}^{i(o)}({\bf x})$ for electric (E)
multipoles where $q=\sqrt{(\omega/c)^2 -k_z^2}$ and incoming (i) and
outgoing (o) waves differ in the definition of the vector fields ${\bf
  V}^i({\bf x}) = \hat {\bf z} J_m(q r) e^{im\phi} e^{ik_z z}$, ${\bf
  V}^o({\bf x}) = \hat {\bf z} H^{(1)}_m(q r) e^{im\phi} e^{ik_z z}$.
Here $(r,\phi,z)$ denote cylindrical coordinates and $J_m$,
$H_m^{(1)}$ are Bessel and Hankel functions of the first kind. In this
basis the matrices of Eq.~(\ref{eq:EM-def-matrix-M}) assume a simple
form where the multi-index $X$ represent now the polarization (M or
E), the wave vector $k_z$ along the cylinder axis and the partial wave
index $m$. The T-matrix is diagonal in polarization, $k_z$ and $m$
with diagonal elements
\begin{eqnarray}
  \label{eq:T_perf_metal_WR_mm}
  T_{M k_z m} &= & (-1)^m \frac{\pi}{2i} \frac{I_m'(qR)}{ K'_m(qR)} \\
  \label{eq:T_perf_metal_WR_ee}
  T_{E k_z m} &= & (-1)^m \frac{\pi}{2i} \frac{I_m(qR)}{ K_m(qR)} \, ,
\end{eqnarray}
where we have applied a Wick rotation $\omega=ic\kappa$ which leads to
modified Bessel functions of the first ($I_n$) and second ($K_n$)
kind. The translation matrices are diagonal in polarization and
$k_z$, and the elements are identical for both polarizations,
\begin{eqnarray}
  \label{eq:B-matrix-uni-WR}
  U_{12,M k_z nm} &=& U_{12,E k_z nm} = \frac{2}{i\pi} 
(-i)^{m-n} K_{m-n} (pd) \\ 
U_{21,M k_z nm} &=& U_{21,E k_z nm} = \frac{2}{i\pi} 
i^{m-n} K_{m-n} (pd) 
\end{eqnarray}
with $p=\sqrt{\kappa^2+k_z^2}$. Due to the decoupling of electric and
magnetic multipoles, corresponding to transverse magnetic (TM) and
transverse electric (TE) field modes, respectively, the Casimir energy of
Eq.~(\ref{eq:energy_gen}) has two independent contributions, ${\cal
  E}_2={\cal E}_{\rm TM}+{\cal E}_{\rm TE}$, with
\begin{equation}
  \label{eq:energy_2D_in}
  {\cal E}_{\rm TM(TE)} = \frac{\hbar c L}{4\pi} \int_0^\infty p d p 
\ln \det (1 - \mathbb{N}_{\rm TM(TE)} ) \, ,
\end{equation}
where $L$ ($\to \infty$) is the cylinder length and the determinant runs
here only over the partial wave indices $m$, $m'$ of the matrix elements
\begin{eqnarray}
  \label{eq:matrix_N_elements}
  N_{TE,mm'} = i^{m'-m} \sum_n 
\frac{I'_m(pR)}{K'_m(pR)} K_{n+m}(pd) \frac{I'_n(pR)}{K'_n(pR)} K_{m'+n}(pd) \\
 N_{TM,mm'} = i^{m'-m} \sum_n
\frac{I_m(pR)}{K_m(pR)} K_{n+m}(pd) \frac{I_n(pR)}{K_n(pR)} K_{m'+n}(pd) \, .
\end{eqnarray}
This result for the energy can be also obtained from a scalar field
theory where TM (TE) modes correspond to Dirichlet (Neumann) boundary
conditions \cite{Rahi:2008kb,Emig:2006a}. 

At large separations $d\gg R$, only matrix elements with $m=m'=0$ for
TM modes and $m=m'=0,\pm 1$ for TE modes contribute to the energy.
When the determinant in Eq.~(\ref{eq:energy_2D_in}) is restricted to
these elements, we find for the interaction of two cylinders for large
$d/R$ to leading order
\begin{equation}
\label{eq:E-two-cylinders}
\begin{split}
{\cal E}_{\rm TM} & = -\hbar c L \frac{1}{8\pi} \frac{1}{d^2 \ln^2(d/R))} 
\left(1-\frac{2}{\ln(d/R)} + \ldots \right) \, , \\
{\cal E}_{\rm TE} & = -\hbar c L \frac{7}{5\pi}\frac{R^4}{d^6} \, .
\end{split}
\end{equation}
The asymptotic interaction is dominated by the contribution from TM modes
that vanishes for $R\to 0$ only logarithmically. 

For arbitrary separations higher order partial waves have to be
considered. The number of partial waves has to be increased with
decreasing separation. A numerical evaluation of the determinant and
integration has revealed an exponentially fast convergence of the
energy in the truncation order for the partial waves, leading to the
results shown in Fig.~\ref{Fig:2cylinders} \cite{Rahi:2008kb}.  It
should be noted that the minimum in the curve for the total
electromagnetic energy results from the scaling by the PFA estimate of
the energy. The total energy is monotonic and the force attractive at
all separations. 

The interaction between cylinders is very distinct from the Casimir or
van der Waals interaction which is reported in literature
\cite{Parsegian:2005eu}. Usually, the interaction is proportional to
the volumes of the interacting objects, i.e., for two spheres of
radius $R$ where the Casimir energy $\sim R^6/d^7$. This scaling with
volumes follows also from a pairwise summation of two-body forces.
However, from the interaction of two parallel plates one knows that
the interaction can scale also with the surface area. This two
examples would suggest for two parallel cylinders of length $L$ an
interaction energy $\sim L R^4/d^6$ or $\sim L R^2/d^4$. But the
actual results of Eq.~(\ref{eq:E-two-cylinders}) has a much weaker,
only logarithmic dependence on the radius. Is is interesting to look
at the variation of the decay exponent of $d$ for the Casimir energy
as a function of the codimension of the object. The exponent is
$(-3,-2+\epsilon,-7)$ for codimensions $1$ (plates), $2$ (cylinders),
$3$ (spheres), respectively, and hence {\it not monotonic}. For a
codimension of two, the Casimir interaction is most long ranged.  The
physical reason for the unexpected scaling of the cylinder interaction
is explained by considering spontaneous charge fluctuations.  On a
sphere, the positive and negative charges can be separated by at most
distances of order $R\ll d$.  The retarded van der Waals interactions
between the dipoles on the spheres lead to the Casimir--Polder
interaction \cite{Casimir:1948a}.  In the cylinder, fluctuations of
charge along the axis of the cylinder can create arbitrary large
positively (or negatively) charged regions.  The retarded interaction
of these charges (not dipoles) gives the dominant term of the Casimir
force.  This interpretation is consistent with the difference between
the two types of polarizations, since for TE modes such charge
modulations cannot occur due to the absence of an electric field along
the cylinder axis, cf.~Eq.~(\ref{eq:E-two-cylinders}) and
Fig.~\ref{Fig:2cylinders}.

\begin{figure}[h]
\includegraphics[width=8cm]{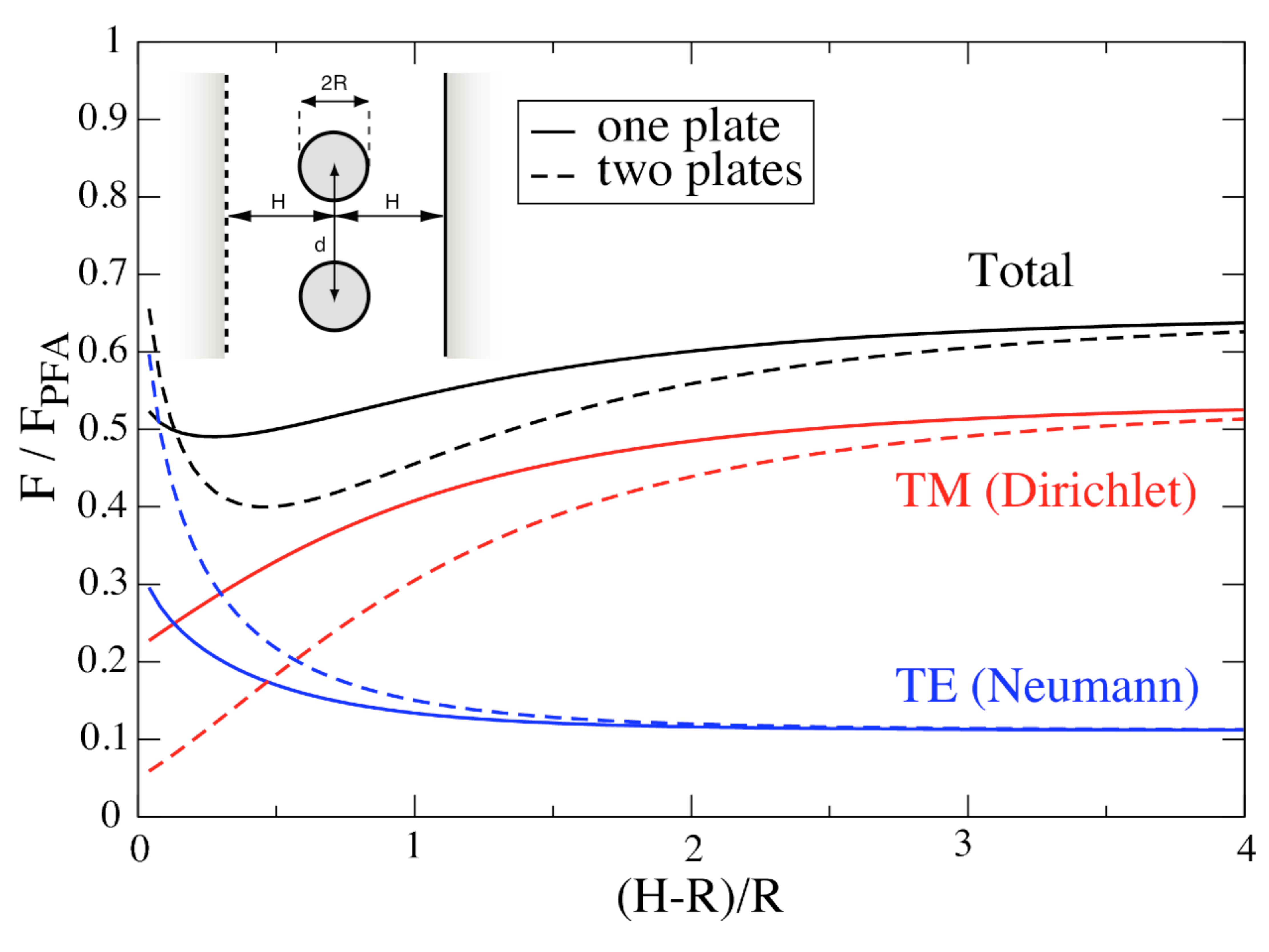}
\caption{\label{Fig:2cylinders+plates}Casimir force between two
  cylinders parallel to one plate or sandwiched between two plates vs.
  the ratio of sidewall separation to cylinder radius $(H-R)/R$, at
  fixed distance $d=4R$ between the cylinders, normalized by the total
  PFA force per unit length between two isolated cylinders,
  $F_{\rm PFA}= -\frac{5}{2}(\hbar c\pi^3/
  1920)\sqrt{R/(d-2R)^7}$. The solid lines
  refer to the case with one plate, while dashed lines depict the
  results for two plates.  Also shown are the individual TE (blue)
  and TM (red) contributions to the force.}
\end{figure}

As a second example, the effect of sidewalls on the interaction of two
cylinders is considered. The geometry consisting of either one or two
plates at a separation $H$ from the two cylinders is shown in
Fig.~\ref{Fig:2cylinders+plates}. For this type of geometry the mean
stress tensor has been computed numerically and it has been observed
that the force between two one-dimensional structures changes
nonmonotonically when $H$ is increased
\cite{Rodriguez:2007a,Rahi:2008bv}. This many-body effect can be
studied by the scattering approach. Instead of studying directly the
interaction of the cylinders and plates via their T-matrices, it is
more convenient to employ the method of images to describe the effect
of the sidewalls \cite{Emig:2008ee,Rahi:2008kb}. For perfectly
conducting sidewalls, their effect on the electromagnetic field can be
taken into account by replacing the free space Green's function by a
half-space or slab Green's function. This results in an expression for
the Casimir energy similar to
Eq.~(\ref{eq:EM-Casimir-energy-gen-result}) that depends only on the
T-matrices of the two cylinders and translation matrices that connect
the original cylinders and their mirror images. The expression of the
energy can be computed again numerically by truncating the partial
wave expansion at sufficiently high order. The resulting Casimir force
between two cylinders with one or two sidewalls as function of the
sidewall separation $H$ is shown in Fig.~\ref{Fig:2cylinders+plates}.
Two interesting features can be observed. First, the attractive total
force varies non-monotonically with $H$: Decreasing for small $H$ and
then increasing towards the asymptotic limit between two isolated
cylinders for large $H$, cf. Eq.~(\ref{eq:E-two-cylinders}). The
extremum for the one-sidewall case occurs at $H-R \approx 0.27 R$, and
for the two-sidewall case is at $H-R \approx 0.46 R$. Second, the
total force between the cylinders for the two-sidewall case in the
proximity limit $H\to R$ is larger than for $H/R \to\infty$. As might
be expected, the $H$-dependence for one sidewall is weaker than for
two sidewalls, and the effects of the two sidewalls are not additive:
not only is the difference from the $H\to\infty$ force not doubled for
two sidewalls compared to one, but the two curves actually intersect.

A simple generic argument for the non-monotonic sidewall effect has
been given in \cite{Rahi:2008bv}. It arises from a competition between
the force from TE and TM polarizations as demonstrated by the results
in Fig.~\ref{Fig:2cylinders+plates}. An intuitive perspective for the
qualitatively different behavior of the TE and TM force as a function
of the sidewall distance is obtained from the method of images.  For
the TM polarization (corresponding to Dirichlet boundary conditions in
a scalar field theory) the Green's function is obtained by subtracting
the contribution from the image so that the image sources have {\it
  opposite} signs.  Any configuration of fluctuating TM charges on one
cylinder is thus screened by images, more so as $H$ is decreased, {\it
  reducing} the force on the fluctuating charges of the second
cylinder.  This is similar to the effect of a nearby grounded plate on
the force between two opposite electrostatic charges. Since the
reduction in force is present for every charge configuration, it is
there also for the average over all configurations.

By contrast, the TE polarization (corresponding to Neumann boundary
conditions in a scalar field theory) requires image sources of the {\it
  same} sign. The total force between fluctuating sources on the
cylinders is now larger and increases as the plate separation $H$ is
reduced.  Note, however, that while for each fluctuating source
configuration, the effect of images is additive, this is not the case
for the average over all configurations. More precisely, the effect of
an image source on the Green's function is not additive because of
feedback effects: the image currents change the surface current
distribution, which changes the image, and so forth.  For example, the
net effect of the plate on the Casimir TE force {\it is not} to double
the force as $H\to R$.  The increase is in fact larger than two due to
the correlated fluctuations.

A similar but weaker non-monotonic dependence on $H$ of the force
between the cylinders is also observed for separations $d$ that are
different from the particular choice in
Fig.~\ref{Fig:2cylinders+plates}. Also, the force between the
cylinders {\it and the sidewalls} is not monotonic in $d$ but the
non-monotonicity is then smaller since the effect of a cylinder on
the force between two bodies is smaller than the effect of an infinite
plate.

\subsection{Spheres}

So far geometries with a direction of translational invariance have
been considered. In the limit of ideal metal surfaces, this invariance
lead to a decoupling of the two polarizations of the electromagnetic
field. Any geometry of experimental interest will
obviously lack this symmetry beyond some length scale. Hence, it is
important to study geometries without this symmetry. {\it
  Compact} objects of arbitrary shape obviously do not have an
invariant direction. Therefore, the two polarisations are coupled and
the matrices in Eq.~(\ref{eq:EM-def-matrix-M}) assume a more
complicated form. A natural choice for a basis are now vector
spherical waves for which the translation matrices ${\mathbb
  U}_{\alpha\alpha',XX'}$ carry an index $X=(E \,{\rm or}\, M,l,m)$
which represents polarisation E or M and the order $l\ge 1$,
$m=-l,\ldots,l$ of the spherical waves. In contrast to the cylindrical
matrices of Eq.~(\ref{eq:B-matrix-uni-WR}), the translation matrix
couples now E and M polarisation and all matrix elements are known
explicitly \cite{Emig:2008ee}. 

\begin{figure}[h]
\includegraphics[width=8cm]{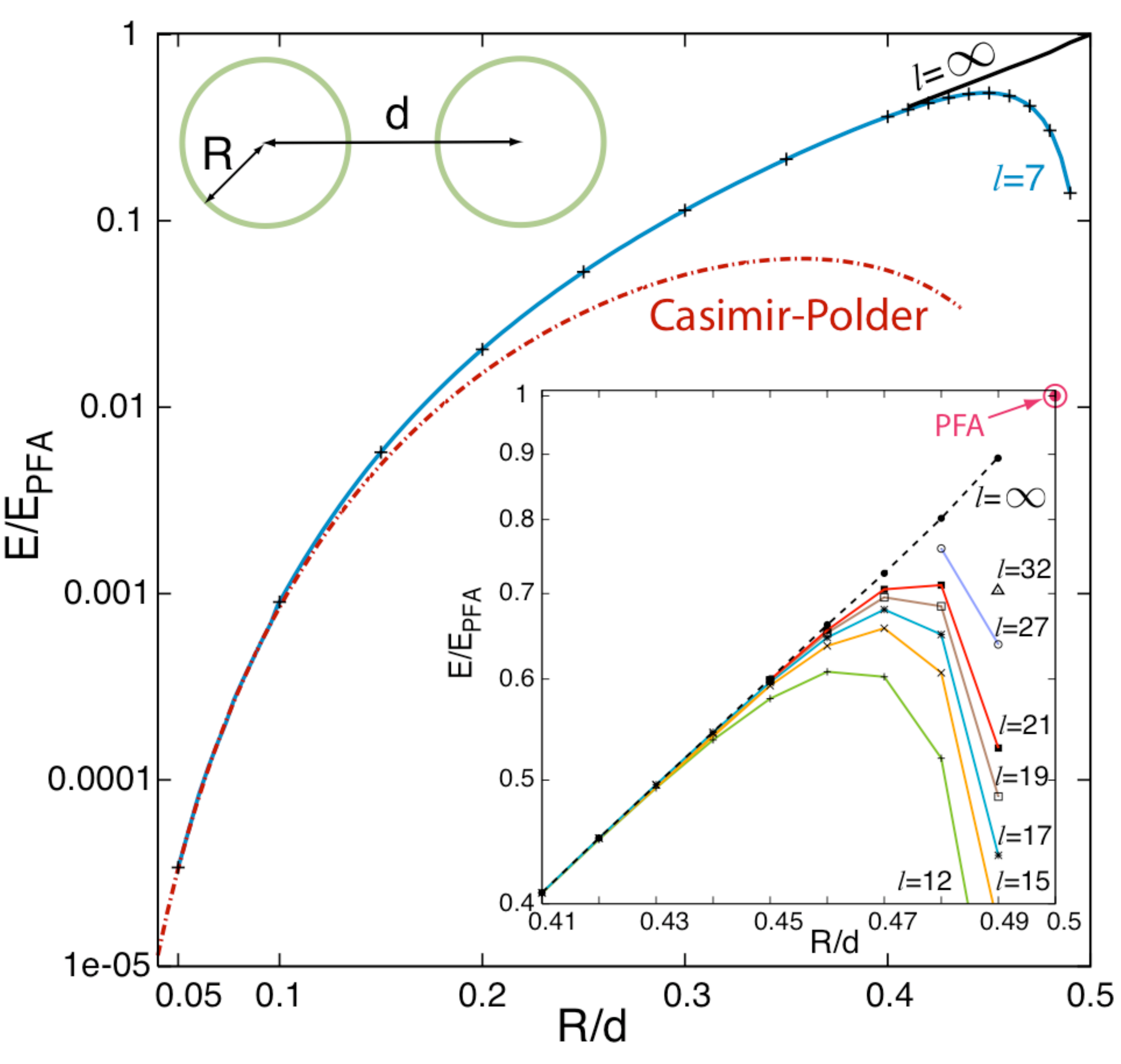}
\caption{\label{Fig:2spheres} Casimir energy of two metal spheres,
  divided by the PFA estimate ${\cal E}_{\rm PFA}=-(\pi^3/1440)\hbar c
  R/(d-2R)^2$, which holds only in the limit $R/d\to 1/2$. The label
  $l$ denotes the multipole order of truncation. The curves $l=\infty$
  are obtained by extrapolation. The Casimir-Polder curve is the
  leading term of Eq.~(\ref{eq:2-metal-spheres}). Inset: Convergence
  with the truncation order $l$ for partial waves at short
  separations.}
\end{figure}

Here we focus on the simplest case of two compact objects: two perfect
metal spheres of equal radius $R$ and center-to-center separation $d$,
see Fig.~\ref{Fig:2spheres}. The T-matrix of a dielectric sphere is
known from the Mie theory for scattering of electromagnetic waves from
spherical particles. Due to spherical symmetry, the E and M
polarisations for all $l$, $m$ are decoupled so that the T-matrix is
diagonal and the coupling of polarisations occurs through the
translation matrices only.  After a Wick rotation to imaginary
frequency $\omega=ic\kappa$ the matrix elements assume in the perfect
metal limit the form
\begin{eqnarray}
  \label{eq:t-matrix-elem-cond-sphere-m}
   T_{MM\,lml'm'}&=& (-1)^l \frac{\pi}{2} \frac{I_{l+{1\over 2}}(\kappa R)}
{K_{l+{1\over 2}}(\kappa R)}
\delta_{ll'}\delta_{mm'}\\
  \label{eq:t-matrix-elem-cond-sphere-e}
T_{EE\,lml'm'}&=& (-1)^l \frac{\pi}{2} \frac{I_{l+{1\over 2}}(\kappa R)+2\kappa R I'_{l+{1\over 2}}(\kappa R)}
{K_{l+{1\over 2}}(\kappa R)+2\kappa R K'_{l+{1\over 2}}(\kappa R)} \delta_{ll'}\delta_{mm'}\, .
\end{eqnarray}
Substitution of these matrix elements together with those of ${\mathbb
  U}_{\alpha\alpha'}$ from \cite{Emig:2008ee} in
Eq.~(\ref{eq:energy_gen}) yields the Casimir energy of two spheres.
For asymptotically large $d$, the energy has only contributions from
$l=l'=1$ (dipoles) and one obtains the Casimir-Polder interaction
between two polarizable particles \cite{Casimir:1948a} where the
electric and magnetic dipole polarizabilities of a perfect metal
sphere are given by $\alpha_E=R^3$ and $\alpha_M=-R^3/2$. This result
can be extended to smaller separations by including higher order
multipoles with $l>1$ that generate higher powers of $R/d$. One
obtains the asymptotic series \cite{Emig:2007os}
\begin{equation}
  \label{eq:2-metal-spheres}
  {\cal E}_2=-\frac{\hbar c}{\pi} \frac{R^6}{d^7} \sum_{n=0}^\infty c_n \left(\frac{R}{d}\right)^n \, ,
\end{equation}
where the first 8 coefficients are $c_0\!\!\!=\!\!\!143/16$,
$c_1\!\!\!=\!\!\!0$, $c_2\!\!\!=\!\!\!7947/160$,
$c_3\!\!\!=\!\!\!2065/32$, $c_4\!\!\!=\!\!\!27705347/100800$,
$c_5=-55251/64$, $c_6=1373212550401/144506880$, $c_7=-7583389/320$.
The energy at all separations can be obtained by truncating the matrix
${\mathbb N}$ defined below Eq.~(\ref{eq:energy_gen}) at a finite
multipole order $l$, and by computing the determinant and the integral
numerically. The result is shown in Fig.~\ref{Fig:2spheres}.  It
provides the force for all separations between the Casimir-Polder
limit for $d\gg R$, and the PFA result for $R/d \to 1/2$.  At a
surface-to-surface distance $4R/3$ ($R/d=0.3$), the PFA overestimates
the energy by a factor of 10. Including up to $l=32$ partial wave
orders and extrapolating based on an exponential convergence in $l$,
the Casimir energy has been determined down to $R/d=0.49$
\cite{Emig:2007os}. The interaction between a sphere and a plate has
been obtained recently and deviations from the PFA have been
quantified \cite{Emig:2008ee}.

\newpage 

\section{Dependence on material properties}

In the previous Sections we have considered perfectly conducting
bodies. For real metals with finite, frequency dependent
conductivity, or more general dielectric media, Casimir interactions
are modified. This is a natural consequence of the fact that dipole
and higher multipole polarizabilities depend on the material
properties of a body. Thus, the induced fluctuating currents depend
not only on shape but also on material composition. For practical
applications and experimental tests it is therefore important to
understand the collective effects of shape and material on Casimir
interactions.  A macroscopic theory that fully accounts for the
material dependence of the interaction between two {\it planar}
surfaces has been established by Lifshitz \cite{Lifshitz:1956ud} in
1956. Until recently, only approximations, limited to short
separations between bodies or sufficiently diluted media, have been
available to study the interaction of dielectric media of arbitrary
shapes. The scattering approach described in the previous Section has
paved the way for studying in detail the material and shape dependence
of Casimir forces beyond the case of planar surfaces and without the
common approximations. We first provide a simple derivation of the
Lifshitz result for two surfaces within the scattering approach. Then
we focus on an example that is of particular interest to the behavior
of nano-particles as they appear, i.e., in suspensions where correlations
between material and shape effects are important.

\subsection{Lifshitz formula}

Consider two material halfspaces that are bounded by planar, parallel
surfaces with a vacuum gap of width $d$ between them. The material in
the two halfspaces can be different and is characterized by the
dielectric functions $\epsilon_\alpha(\omega)$ and magnetic
permeabilities $\mu_\alpha(\omega)$ where $\alpha=1,2$ numbers the
halfspaces.  A compact derivation of the Casimir-Lifshitz interaction
between the two surfaces follows from the scattering formula of
Eq.~(\ref{eq:EM-Casimir-energy-gen-result}). The T-matrix of a planar
dielectric surface is given by the Fresnel coefficients which are
usually expressed in a planar wave basis. When we define the two
polarizations relative to the surface normal vector, the T-matrix is
diagonal in polarization and in the wave vector $k_\|$ parallel to
the surface. The diagonal matrix elements are
\begin{equation}
\begin{split}
  \label{eq:T_planar}
  T_{\alpha,M\,k_\|} &= \frac{\mu_\alpha(ic\kappa) p -p_\alpha}{\mu_\alpha(ic\kappa) p+p_\alpha} , \\
  T_{\alpha,E\,k_\|} &= \frac{\epsilon_\alpha(ic\kappa) p -p_\alpha}{\epsilon_\alpha(ic\kappa) p+p_\alpha} 
\end{split}
\end{equation}
where $p=\sqrt{\kappa^2+k_\|^2}$ and
$p_\alpha=\sqrt{\epsilon_\alpha(ic\kappa)\mu_\alpha(ic\kappa)
  \kappa^2+k_\|^2}$.  The translation matrices for translations
perpendicular to the surfaces by a distance $d$ are also diagonal in
$k_\|$ in the planar wave basis and the diagonal elements have the
simple form
\begin{equation}
  \label{eq:U_planar}
U_{\alpha\alpha',M\, k_\|} = U_{\alpha\alpha',E\, k_\|}  = e^{-p d} 
\end{equation}
for $\alpha\neq\alpha'=1,2$, i.e., they do not couple E and M
polarizations and are identical for the two polarizations.  The
determinant of Eq.~(\ref{eq:EM-Casimir-energy-gen-result}) leads to a
product over all $k_\|$ which becomes an integral after taking the
logarithm. The resulting Casimir-Lifshitz energy has two separate
contributions form M and E polarizations (TE and TM modes,
respectively),
\begin{eqnarray}
  \label{eq:E_Casimir_Lifshitz}
  {\cal E} &=& \frac{\hbar c A}{4\pi^2} \int_0^\infty d\kappa \int_0^\infty
\, k_\| dk_\| \ln\left[\left(1-\frac{\epsilon_1(ic\kappa)p-p_1}{\epsilon_1(ic\kappa)p+p_1}
\frac{\epsilon_2(ic\kappa)p-p_2}{\epsilon_2(ic\kappa) p+p_2}e^{-2pd}\right) \right.
\nonumber\\
&& \left. \times \left(1-\frac{\mu_1(ic\kappa)p-p_1}{\mu_1(ic\kappa)p+p_1}
\frac{\mu_2(ic\kappa)p-p_2}{\mu_2(ic\kappa) p+p_2}e^{-2pd}\right)
\right] \, ,
\end{eqnarray}
where $A$ is the surface area. This results generalizes the Casimir
interaction between two perfect metal plates of
Eq.~(\ref{eq:Casimir_E_plates}) to dielectric materials.

\subsection{Nanoparticles: quantum size effects}

The Lifshitz formula of Eq.~(\ref{eq:E_Casimir_Lifshitz}), while
derived for infinitely extended planar surfaces, is commonly applied
within a proximity approximation also to curved surfaces of particles
of finite size. This leads to predictions for the interaction that are
limited to particles that a very large compared to their separations.
To be able to study the interaction of particles of arbitrary sizes
and separations, a theory is needed that is a generalization of the
Lifshitz formula to bodies of arbitrary shape. Such a general theory
provides the scattering formula of
Eq.~(\ref{eq:EM-Casimir-energy-gen-result}). The challenges in
applying this formula consist in the computation of the T-matrix for
bodies with general dielectric functions and in the proper
modelling of the dielectric response of the bodies. The latter is
especially important for nano-particles for which  bulk
optical properties are modified by finite-size effects. 

Some characteristic effects of the Casimir interaction between
nano-particles will be discussed in this Section by studying two
spheres with {\it finite} conductivity in the limit where their radius
$R$ is much smaller than their separation $d$. We assume further that
$R$ is large compared to the inverse Fermi wave vector $\pi/k_F$ of
the metal. Since typically $\pi/k_F$ is of the order of a few
Angstrom, this assumption is reasonable even for nano-particles. To
employ Eq.~(\ref{eq:EM-Casimir-energy-gen-result}), we need the
T-matrix of a sphere with general dielectric function
$\epsilon(\omega)$ which generalizes the matrix of
Eqs.~(\ref{eq:t-matrix-elem-cond-sphere-m}),
(\ref{eq:t-matrix-elem-cond-sphere-e}). All elements of this matrix
are known explicitly, see, e.g., \cite{Emig:2007os}. Relevant to the
interaction for $d\gg R$ are the dipole matrix elements ($l=l'=1$) at
low frequencies $\kappa$. To proceed, we need information about the
dielectric function on the imaginary frequency axis $\omega=ic\kappa$
for small $\kappa$. Theories for the optical properties of small
metallic particles \cite{Wood:1982nl} suggest a Drude like response
\begin{equation}
  \label{eq:eps_Drude}
  \epsilon(ic\kappa) = 1+4\pi \frac{\sigma(ic\kappa)}{c\kappa} \, ,
\end{equation}
where $\sigma(ic\kappa)$ is the conductivity which approaches for
$\kappa\to 0$ the dc conductivity $\sigma_{dc}$. For bulk metals
$\sigma_{dc}=\omega_p^2\tau/4\pi$ where $\omega_p=\sqrt{4e^2k_F^3/3\pi
  m_e}$ is the plasma frequency with electron charge $e$ and electron
mass $m_e$, and $\tau$ is the relaxation time. With decreasing
dimension of the particle, $\sigma_{dc}(R)$ is reduced compared to its
bulk value due to finite size effects and hence becomes a function of
$R$ \cite{Wood:1982nl}. 

In the low frequency limit, with $\epsilon(ic\kappa)$ of
Eq.~(\ref{eq:eps_Drude}) the T-matrix elements for magnetic and
electric dipole scattering $(l=l'=1$) are diagonal in $m$ and have the
series expansion
\begin{eqnarray}
  \label{eq:T-Drude-M}
  T_{MM\,1m1m} &=& -\frac{4\pi}{45} \frac{R\sigma_{dc}(R)}{c}
(\kappa R)^4 + \ldots \\
  \label{eq:T-Drude-E}
T_{EE\,1m1m} &=& \frac{2}{3} (\kappa R)^3 -\frac{1}{2\pi} 
\frac{c}{R\sigma_{dc}(R)} (\kappa R)^4 +\ldots \, .
\end{eqnarray}
To leading order $\sim \kappa^3$ the electric dipole matrix elements
are identical to those of a perfectly conducting sphere and finite
conductivity modifies higher orders only. In the magnetic dipole
matrix elements, however, the leading term $-(\kappa R)^3/3$ of the
perfect conductor result of Eq.~(\ref{eq:t-matrix-elem-cond-sphere-m})
is absent. This is consistent with the observation that the magnetic
dipole polarizability is reduced by a factor $\sim (\kappa R)^2 [\epsilon(ic\kappa)-1]$ and $\epsilon(ic\kappa)-1\sim \kappa^{-1}$ due
to Eq.~(\ref{eq:eps_Drude}). 

When the matrix elements of Eqs.~(\ref{eq:T-Drude-M}),
(\ref{eq:T-Drude-E}) together with the translation matrices ${\mathbb
  U}_{\alpha\alpha'}$ in spherical coordinates are substituted into
Eq.~(\ref{eq:energy_gen}) an expansion for large distance $d$ yields
the Casimir energy of two spheres
\begin{equation}
  \label{eq:E_drude}
\frac{E}{\hbar c} = -\frac{23}{4\pi} \frac{R^6}{d^7} -\left(
\frac{R\sigma_{dc}(R)}{c} -\frac{45}{4\pi^2} \frac{c}{R\sigma_{dc}(R)}\right) \frac{R^7}{d^8} + \ldots \, .   
\end{equation}
The leading term is material independent but different from that of
the perfect metal sphere interaction of Eq.~(\ref{eq:2-metal-spheres})
since only the electric polarization contributes to it. At next order,
the first and second terms in the parentheses come from magnetic and
electric dipole fluctuations, respectively. Notice that the term $\sim
1/d^8$ is absent in the interaction between perfectly conducting
spheres, see Eq.~(\ref{eq:2-metal-spheres}). The limit of perfect
conductivity, $\sigma_{dc}\to\infty$ cannot be taken in
Eq.~(\ref{eq:E_drude}) since this limit does not commute with the
low $\kappa$ or large $d$ expansion.

\begin{figure}[h]
\includegraphics[width=11cm]{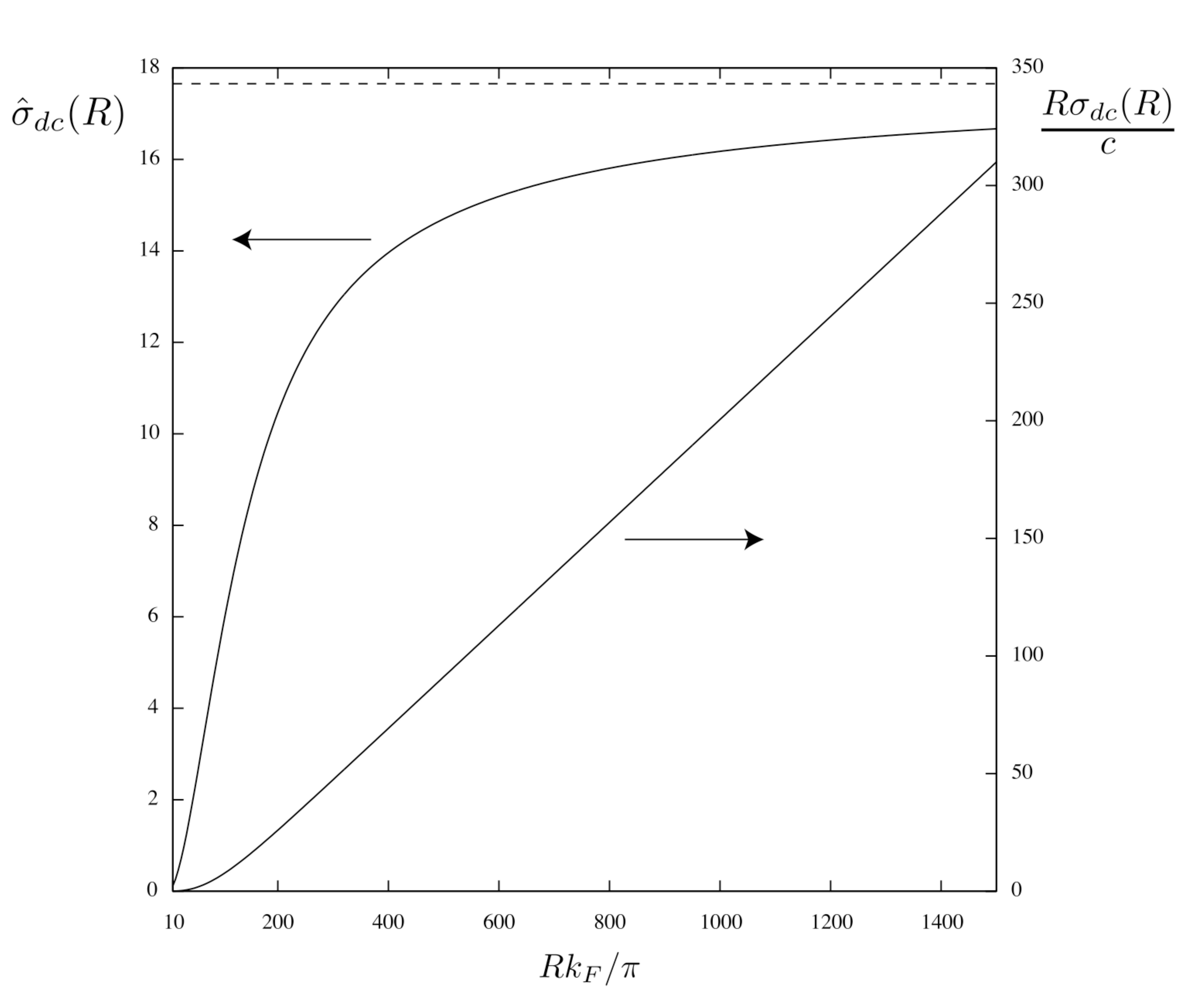}
\caption{\label{Fig:sigma} Dimensionless dc conductivity
  $\hat\sigma_{dc}(R)$ in units of $e^2/2\hbar a_0$ (with Bohr radius
  $a_0$) for a Aluminum sphere with $\epsilon_F=11.63$eV,
  $\pi/k_F=1.8\!\!\buildrel _\circ \over {\mathrm{A}}$ and
  $\tau=0.8\cdot 10^{-14}$sec as function of the radius $R$, measured
  in units of $\pi/k_F$, see Eq.~(\ref{eq:sigma_dc}). Also shown is
  the corresponding ratio $R\sigma_{dc}(R)/c$ that determines the
  Casimir interaction of Eq.~(\ref{eq:E_drude}). The bulk dc
  conductivity $\hat\sigma_{dc}(\infty)=17.66$ is indicated by the
  dashed line.}
\end{figure}

In order to estimate the effect of finite conductivity and its
dependence on the size of the nano-particle, we have to employ a
theory that can describe the evolution of $\sigma_{dc}(R)$ with the
particle size. A theory for the dielectric function of a cubical
metallic particle of dimensions $R \gg \pi/k_F$ has been developed
within the random phase approximation in the limit of low frequencies
$\ll c/R$ \cite{Wood:1982nl}. In this theory it is further assumed
that the discreteness of the electronic energy levels, and not the
inhomogeneity of the charge distribution, is important. This implies
that the particle responds only at the wave vector of the incident field
which is a rather common approximation for small particles.
From an electron number-conserving relaxation time approximation
the complex dielectric function is obtained which yields the 
size-dependent dc conductivity  for a cubic particle
of volume $a^3$ \cite{Wood:1982nl}. It has been shown that 
the detailed shape of the particle does not matter much, and we can
set $a=(4\pi/3)^{1/3}R$ which defines the volume equivalent sphere
radius $R$. This yields the estimate
\begin{eqnarray}
  \label{eq:sigma_dc}
  \sigma_{dc}(R) &=& \sigma_{dc}(\infty) \left[
1-\frac{3\pi k_F a+\pi^2}{4(k_F a)^2}
-\frac{48\pi}{(k_Fa)^3\Gamma^2}\right.\\
&\times&\left.\!\!\!{\rm Re} \sum_{m=1}^{k_Fa/\pi} m^2((k_Fa/\pi)^2-m^2) \times \left\{
\begin{array}{lr}
-z_m \tan z_m & m \,{\rm even}\\
+z_m \cot z_m & m \, {\rm odd}
\end{array}
\right. 
\right] \nonumber
\end{eqnarray}
with
\begin{equation}
  \label{eq:z_m}
  z_m=\frac{\pi m}{2} \sqrt{1-\frac{i\Gamma}{m^2}}
\end{equation}
and where $\sigma_{dc}(\infty)=\omega_p^2\tau/4\pi$ is the bulk Drude
dc conductivity and $\Gamma=(\hbar/\tau \epsilon_F)(k_F a/\pi)^2$ is a
linewidth with Fermi energy $\epsilon_F$. The factor in square
parentheses multiplying $\sigma_{dc}(\infty)$ describes quantum size
effects and leads to a substantial {\it reduction} of the dc
conductivity for nano-scale particles. While the above expression is
applicable for $\pi/k_F \ll a$, it suggest that for $\pi/k_F\simeq a$
the particle ceases to be conducting which is consistent with a
metal-insulator transition due to the localisation of electrons for
particles with a size of the order of the mean free path. It is
instructive to consider the size dependence of $\sigma_{dc}(R)$ and of
the Casimir interaction for a particular choice of material. Following
\cite{Wood:1982nl}, we focus on small Aluminum spheres with
$\epsilon_F=11.63$eV and $\tau=0.8\cdot 10^{-14}$sec. These parameters
correspond to $\pi/k_F=1.8\!\!\buildrel _\circ \over {\mathrm{A}}$ and
a plasma wavelength $\lambda_p=79$nm. It is useful to introduce the
dimensionless conductivity $\hat\sigma_{dc}(R)$, which is measured in
units of $e^2/2\hbar a_0$ with Bohr radius $a_0$, so that the important
quantity of Eq.~(\ref{eq:E_drude}) can be written as
$R\sigma_{dc}(R)/c=(\alpha/2)(R/a_0)\hat\sigma_{dc}(R)$ where $\alpha$
is the fine-structure constant. The results following from
Eq.~(\ref{eq:sigma_dc}) are shown in Fig.~\ref{Fig:sigma}.  For
example, for a sphere of radius $R=10$nm, the dc conductivity is
reduced by a factor $\approx 0.15$ compared to the bulk Drude value.
If the radius of the sphere is equal to the plasma wavelength
$\lambda_p$, the reduction factor $\approx 0.8$. These results show
that shape and material properties are important for the Casimir
interaction between nano-particles. Potential applications include the
interaction between dilute suspensions of metallic nano-particles.

\section{Casimir force driven nanosystems}
\label{sec:driven_system}

We have seen that Casimir forces increase strongly with decreasing
distance and hence it can be expected that they are important in
devices that are composed of moving elements at short separations.
Indeed, a common phenomena seen in nano-mechanical devices is stiction
due to attractive van der Waals and Casimir forces.  This effect
imposes a minimum separation between objects in order to prevent that
they stick together. However, one can make also good use of Casimir
interactions in nano-devices by employing them to actuate components
of small devices without contact
\cite{Ashourvan:2007jy,Emig:2007a,Ashourvan:2007gw}.  In
\cite{Emig:2007a} it has been demonstrated that this can be achieved
by coupling two periodically structured parallel surfaces by the
zero-point fluctuations of the electromagnetic field between them.  We
will consider this effect here as an example for Casimir force induced
non-linear dynamics, providing a direct application of the results
obtained in Section \ref{sec:lateral_force}. We have seen that the
broken translation symmetry parallel to the surfaces results in a
sideways force which has been predicted theoretically
\cite{Emig:2001a,Emig:2003a} and observed experimentally between
static surfaces \cite{Chen:2002a}. If at least one of the surfaces is
structured {\it asymmetrically} there is an additional breaking of
reflection symmetry and the surfaces can in principle be set into
relative lateral motion in the direction of broken symmetry. The
energy for this transport has to be pumped into the system by external
driving. This can be realized by setting the surfaces into relative
oscillatory motion so that their normal distance is an unbiased
periodic function of time. Since the sideways Casimir force decays
exponentially with the normal distance [see Eq.~(\ref{eq:f-pert})],
the surfaces experience an asymmetric periodic potential that varies
strongly in time.

This scenario resembles so-called ratchet systems \cite{Reimann:2002a}
that have been studied extensively during the last decade in the
context of Brownian particles \cite{Ajdari:1997a}, molecular motors
\cite{Astumian:1998a} and vortex physics in superconductors
\cite{Souza-Silva:2006a}, to name a few recent examples. Most of the
works on ratchets consider an external time-dependent driving force
acting on overdamped degrees of freedom to rectify thermal noise. For
nano-systems, however, it has been pointed out that inertia terms due
to finite mass should not be neglected and, actually, can help the
ratchets to perform more efficiently than their overdamped companions
\cite{Marchesoni:2006a}.  Finite inertia typically induce in Langevin
dynamics deterministic chaos that has been shown to be able to mimic
the role of noise and hence to generate directed transport in the
absence of external noise \cite{Mateos:2000a}. Here we use this effect
in the different context of so-called pulsating (or effectively
on-off) ratchets where the strengths of the periodic potential varies
in time \cite{Reimann:2002a}. We consider weak thermal noise only to
test for stability of the inertia induced transport --- not as the
source of driving\footnote{In the absence of inertia, finite
thermal noise {\it is} necessary for on-off ratchets to generate
directed motion.}.

It has been demonstrated that the system described above indeed allows
for directed relative motion of the surfaces due to chaotic dynamics
caused by the lateral Casimir force \cite{Emig:2007a}.  The transport
velocity is stable across sizeable intervals of the amplitude and
frequency of surface distance oscillations and damping. The velocity
scales linear with frequency across these intervals and is almost
constant below a critical mean distance beyond which it drops sharply.
The system exhibits multiple current reversals as function of the
oscillation amplitude, mean distance and damping. This ``Casimir
ratchet'' allows contact-less transmission of motion which is
important since traditional lubrication is not applicable in
nano-devices.  This actuation mechanism should be compared to other
actuation schemes as magnetomotive or capacitive (electrostatic) force
transmission.  The Casimir effect induced actuation has the advantage
of working also for insulators and does not require any electrical
contacts and/or external fields. Other applications of zero-point
fluctuation induced (van der Waals) interactions to nano-devices have
been experimentally realized already to construct ultra-low friction
bearings from multiwall carbon nanotubes \cite{Cumings:2000a}.

%%%
%%% Geometry & Casimir interaction
%%%

\begin{figure}
\includegraphics[width=7cm]{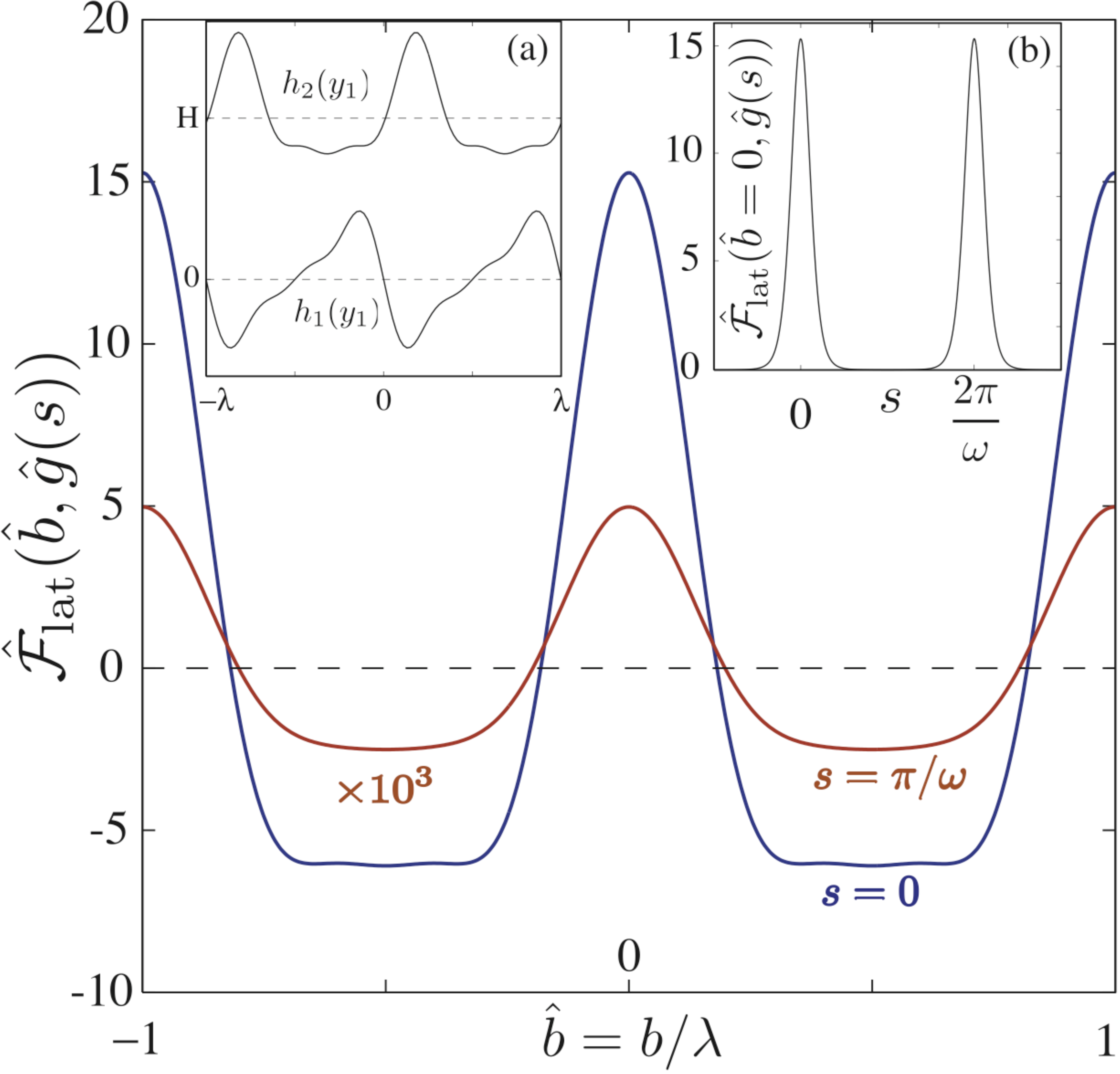}
\caption{\label{Fig:force} The lateral Casimir force acting between
  the two surfaces as function of the shift $\hat b$ at time $s=0$ and half
  period $s=\pi/\omega$ (drawn to a larger scale by a factor $10^3$)
  for parameters $\eta=0.65$, $H_0=0.1 \lambda$.  Insets: (a) Surface
  profiles at their equilibrium position at $\hat b=0.182$ (b)
  Periodic variation of the maximum force at $\hat b=0$ with time.}
\end{figure}

In the following, we consider two (on average) parallel metallic
surfaces with periodic, uni-axial corrugations (along the $y_1$-axis)
that have distance $H$, see inset (a) of Fig.~\ref{Fig:force}. To
begin with, we assume that both surfaces are at rest with a relative
lateral displacement $b$.  Then the surface profiles can be
parametrized as
\begin{eqnarray}
\label{corr-2}
 h_1(y_1) \, &=& \,a \, \sum_{n=1}^\infty c_n
 e^{2\pi i n y_1/\lambda_1} + {\rm c.c.} \, , \\
h_2(y_1) \, &=&  \, a \,\sum_{n=1}^{\infty} d_n 
e^{2\pi i n (y_1-b)/\lambda_2} +{\rm c.c.} \, \, ,
\end{eqnarray}
where $a$ is the corrugation amplitude, $\lambda_1$, $\lambda_2$ are
the corrugation wave lengths, $c_n$, $d_n$ are Fourier
coefficients.

The dependence of the Casimir energy ${\cal E}$ on $H$ and $b$ causes
macroscopic forces on the surfaces.  For a varying separation $H$ this
is the normal Casimir attraction between metallic surfaces, modified
by the corrugations. Below we will assume $H=H(t)$ to be a
time-dependent distance that is kept at a fixed oscillation by an
additional external force from clamping to an oscillator. In such
setup, the surfaces can react freely only to the lateral force
component ${\cal F}_{\rm lat}(b,H)=-\partial {\cal E}/\partial b$.
The results of Section \ref{sec:lateral_force} are readily extended to
periodic profiles of arbitrary shapes as described by
Eq.~(\ref{corr-2}). The corrugation lengths have to be commensurate,
$\lambda_1/\lambda_2=p/q$ with integers $p$, $q$ in order to produce a
finite lateral force per surface area. For the purpose of this
example, it is sufficient to consider the case $p=1$. Generalizing the
result of Eq.~(\ref{Ecc}), the lateral ($b$-dependent) part of the
Casimir energy per surface area can then be written as
\begin{equation}
\label{eq:lateral_energy}
{\cal E}(b)= 
\frac{2\hbar c a^2}{H^5} \sum_{n=1}^\infty \left( c_n d_{-nq} 
e^{-2\pi i n b/\lambda_1} + {\rm c.c.} \right) 
J\left( n \frac{H}{\lambda_1} \right)
\end{equation}
to order $a^2$. The exact form of the function $J(x)=J_{\rm
  TM}(x)+J_{\rm TE}(x)$ is given by Eqs.~(\ref{j-tm}), (\ref{j-te}).
For the present purpose it is sufficient to use the simplified
expression
\begin{equation}
\label{eq:J-fct}
J(x)\simeq\frac{\pi^2}{120}\left(
1 + 2\pi x + \gamma  x^2 + 32 x^4 
\right)\, e^{-2\pi x}
\end{equation}
with $\gamma=12.4133$, which is exact for both asymptotically large and
small $x$ and approximates the exact results with sufficient accuracy
for all $x$. (The maximal deviation from the exact result is
$\approx\pm 0.5\%$ around $x=0.5$.) The Casimir potential of
Eq.~(\ref{eq:lateral_energy}) has two interesting properties which are
useful to the construction of a ratchet. First, it decays
exponentially with $H$, and thus can be essentially switched on and
off periodically in time by oscillating $H$. Second, the potential is
not only periodic in $b$ but acquires asymmetry from the surface profiles 
at small $H\ll \lambda$ and an universal symmetric shape for $H\gg \lambda$
since the effect of higher harmonics of the surface profile is exponentially
diminished as discussed in Section \ref{sec:lateral_force}.

%%%
%%% Equation of motion & relevant scales
%%%

The relative surface displacement $b(t)$ can be considered as a
classical degree of freedom with inertia. Its equation of motion is
described by Langevin dynamics of the form
\begin{equation}
\label{eq:langevin}
\rho \ddot b + \gamma\rho \dot b = {\cal F}_{\rm lat}[b, H(t)] + 
\sqrt{2\gamma\rho T} \, \xi(t) \, ,
\end{equation}
where $\rho$ is the mass per surface area, $\gamma$ the friction
coefficient, $T$ the intensity (divided by surface area) of the
Gaussian noise $\xi(t)$ with zero mean and correlations $\langle
\xi(t)\xi(t')\rangle = \delta(t-t')$ so that the Einstein relation is
obeyed. This stochastic term describes ambient noise due to effects of
temperature and pressure. (Additional contributions from thermally
excited photons to the Casimir force can be neglected at surface
distances well below the thermal wavelength $\hbar c/(2 T)$.) The
system is driven by rigid oscillations of one surface so that the
distance $H(t)=H_0 g(t)$ oscillates about the mean distance $H_0$ with
$g(t)=1-\eta\cos(\Omega t)$. For simplicity, we consider now equal
corrugation lengths $\lambda_1=\lambda_2\equiv\lambda$.  We define the
following dimensionless variables: $\hat b=b/\lambda$, $s=t/\tau$ for
lateral lengths and time with the typical time scale
$\tau=(\lambda/a)\sqrt{\rho H_0^5/\hbar c}$ resulting from a balance
between inertia and Casimir force. Hence velocities will be measured
in units of $v_0=\lambda/\tau$. There are five dimensionless
parameters which can be varied independently for fixed surface
profiles: the damping $\hat \gamma=\tau\gamma$, the angular frequency
$\omega=\tau\Omega$, the driving amplitude $\eta$, the scaled mean
distance $H_0/\lambda$ and the noise intensity $\hat T=(T/\hbar
c)(H_0^5/a^2)$. The dimensionless equation of motion for $\hat b(s)$ reads
\begin{equation}
\label{eq:langevin-dimless}
\ddot{\hat b}+\hat \gamma \dot{\hat b} = \hat {\cal F}_{\rm lat}[\hat b,\hat g(s)]
+\sqrt{2\hat\gamma \hat T} \, \hat\xi(s) 
\end{equation}
with the Casimir force 
\begin{equation}
\label{eq:force-dimless}
\hat {\cal F}_{\rm lat}(\hat b, \hat g)=\frac{4\pi}{\hat g^5} 
\sum_{n=1}^\infty f_n \cos(2\pi n \hat b) J\left( n \hat g \frac{H_0}{\lambda} \right) \, ,
\end{equation}
where we have chosen surface profiles with $c_n=i\sqrt{f_n/(2n)}$,
$d_n=\sqrt{f_n/(2n)}$ with real coefficients $f_n$ in
Eq.~(\ref{corr-2}), and $\hat g(s)=1-\eta\cos(\omega s)$.

%%% 
%%% numerical approach & averages
%%%

Directed transport is possible in certain parameter ranges even in the
deterministic case where noise is absent. However, to probe the
robustness of transport, we consider in the following primarily the limit of
weak noise by choosing $\hat T=10^{-3}$. In fact, it has been shown
for underdamped ratchets with time-independent potentials and periodic
driving that even an infinitesimal amount of noise can change the
rectification from chaotic to stable \cite{Marchesoni:2006a}.  To look
for similar generic behavior of the pulsating ratchet, we consider a
specific geometry consisting of a symmetric and a sawtooth-like
surface profile corresponding to three harmonics with $f_1=0.0492$,
$f_2=0.0241$, $f_3=0.0059$ and $f_n=0$ for $n>3$. Inset (a) of
Fig.~\ref{Fig:force} shows these profiles in their stable position
with $\hat b=0.182$ that minimizes the Casimir energy. The resulting
spatial variation of the Casimir force with $\hat b$ is plotted in
Fig.~\ref{Fig:force} for minimal ($s=0$) and maximal ($s=\pi/\omega$)
surface distance with parameters $H_0/\lambda=0.1$, $\eta=0.65$. It
can be clearly seen that the asymmetry is reduced at larger distance
where the variation of the force becomes more sinusoidal. Inset (b)
shows the on-off-like time-dependence of the force amplitude at $\hat
b=0$ due to the oscillating surface distance.

The non-linear equation of motion of Eq.~(\ref{eq:langevin-dimless})
has to be solved numerically. The trajectory $\hat b(s)$ was obtained
from a second order Runge-Kutta algorithm. As initial conditions 
an equidistant distribution over the interval $[-1,1]$ for $\hat
b(0)$ and $\dot{\hat b}(0)=0$ is used. For each set of parameters 
200 different trajectories are calculated from varying initial conditions
and noise, each evolving over $4 \times 10^3$ periods $2\pi/\omega$ so
that transients have decayed. The average velocity $\langle\!\langle v
\rangle\!\rangle$ involves two different averages of $\dot{\hat
  b}(s)$: The first average is over initial conditions and noise for
every time step, then the averaged trajectory is averaged over all
discrete times of the numerical solution. For an efficient directed
transport it is not sufficient to have only a finite average
$\langle\!\langle v \rangle\!\rangle$. To exclude trajectories with a
high number of velocity reversals, the fluctuations about the average
velocity must be small, i.e., the variance $\sigma^2 =
\langle\!\langle v^2 \rangle\!\rangle -\langle\!\langle v
\rangle\!\rangle^2$ must be smaller than
$\langle\!\langle v \rangle\!\rangle^2$.

%%%
%%% results for velocity & fluctuations of it
%%%

\begin{figure}
\includegraphics[width=0.75\linewidth]{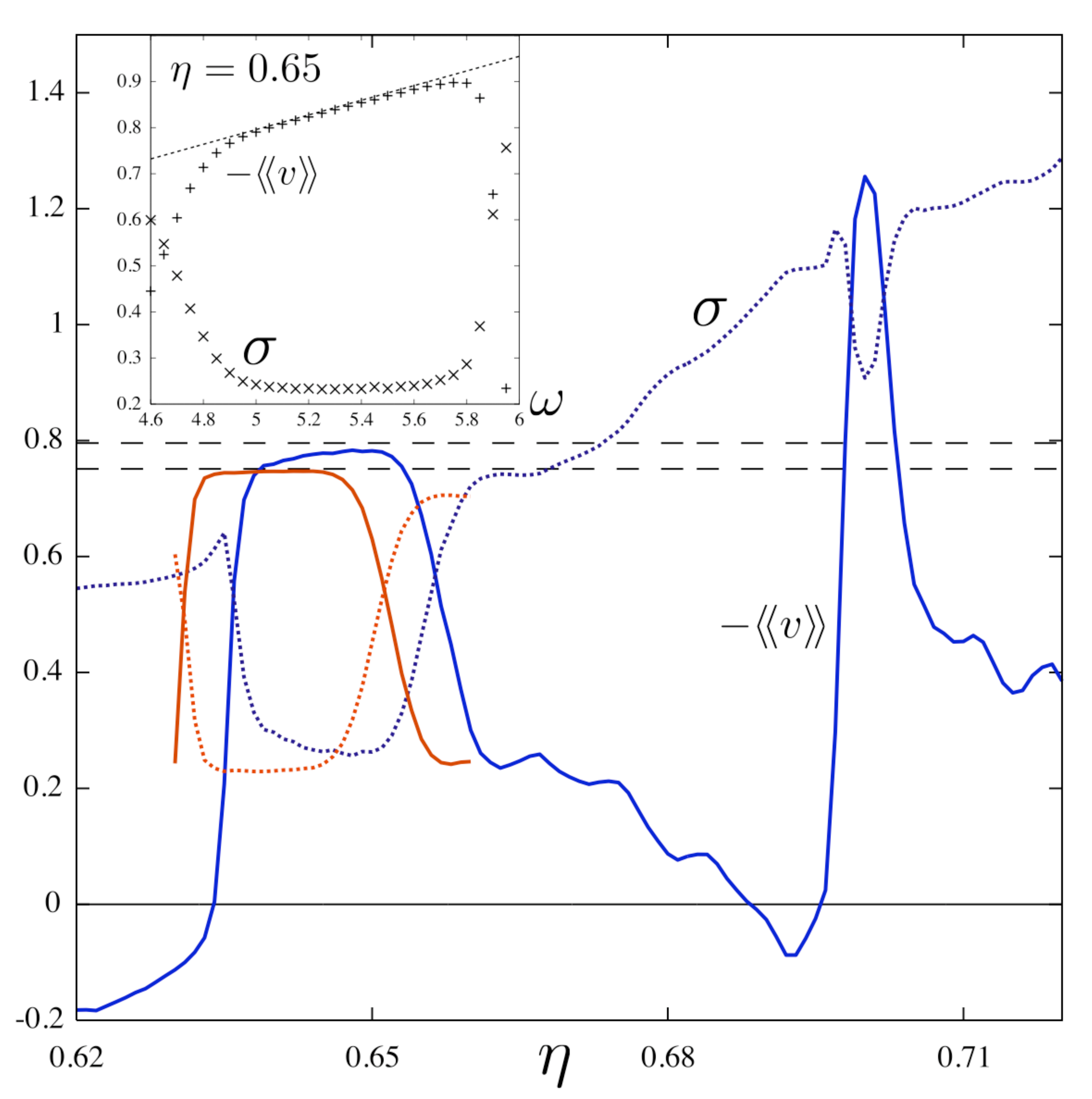}
\caption{\label{Fig:v-of-eta} Mean $\langle\!\langle v
  \rangle\!\rangle$ and standard deviation $\sigma$ of the (negative)
  velocity as function of the driving amplitude $\eta$ for the
  frequencies $\omega=5.0$ and $\omega=4.72$ (for the latter only the
  stable plateau is shown). The parameters are $H_0=0.1\lambda$,
  $\hat\gamma=0.9$, $\hat T=10^{-3}$. Inset: Dependence of the same
  quantities on frequency for fixed $\eta=0.65$. Straight dashed lines
  correspond in both graphs to the velocity $\omega/(2\pi)$.}
\end{figure}

Naively, one can expect directed motion of the surface profile
$h_2(y_1)$ into the positive $y_1$-direction ($\dot{\hat b}<0$) since
the Casimir force in Fig.~\ref{Fig:force} is asymmetric with negative
values lasting for longer time than positive ones. However, the actual
behavior is more complicated due to chaotic
dynamics. Fig.~\ref{Fig:v-of-eta} shows the dependence of the average
velocity and its standard deviation $\sigma$ on the driving amplitude
$\eta$ and frequency $\omega$ for $H_0=0.1 \lambda$, $\hat\gamma=0.9$.
For a fixed frequency there is an optimal interval of driving
amplitudes across which the average velocity is almost constant with
$\langle\!\langle v \rangle\!\rangle \simeq -\omega/(2\pi)$. Small
deviations from the latter value result from noise as has been checked
by studying the dynamics at $\hat T=0$. At higher driving amplitudes
a second narrower interval with maximal $\langle\!\langle v
\rangle\!\rangle$ is observed which is more strongly reduced and smeared out from
its deterministic value $-2\times \omega/(2\pi)$ by noise. At the
plateaus of constant velocity the standard deviation $\sigma$ is
substantially reduced, rendering transport efficient. Outside the
plateaus velocity reversals occur and $\sigma$ increases linearly with
$\eta$. For fixed amplitude $\eta$, the average velocity is stable at
the value $-\omega/(2\pi)$ over a sizeable frequency range (see inset
of Fig.~\ref{Fig:v-of-eta}).

In order to understand the observed behavior it is instructive to
analyze the dynamics in the three dimensional extended phase space.
Attractors of the long-time dynamics can be identified from Poincar\'e
sections using the period $2\pi/\omega$ of the surface oscillation as
stroboscopic time. To obtain a compact section, the trajectory is
folded periodically in $y_1$ on one period of the Casimir potential.
From these sections we can distinguish between periodic and chaotic
orbits. As a start, we consider the deterministic limit with $\hat
T=0$. The plateaus around $\eta=0.65$ and $\eta=0.7$ result both from
periodic orbits of period one, corresponding to a single point in the
Poincar\'e section. On the right (downward) edges of the first
plateaus we observe period doubling, i.e., a periodic attractor
with period two. Upon a further increase of $\eta$, chaotic orbits
dominate the motion. Hence the system exhibits a period-doubling route
to chaos with enhanced velocity fluctuations. The findings apply
basically also to weak noise ($\hat T=10^{-3}$) but the sharp points
of the periodic attractors in the Poincar\'e sections are smeared out
leading to a decreased $\langle\!\langle v \rangle\!\rangle$.  The
transition from chaotic to periodic dynamics at the beginning of the
rising edge of the plateaus is accompanied by a velocity reversal.
This is consistent with the observation for non-pulsating
potentials that velocity reversals are due to a bifurcation from
chaotic to periodic dynamics \cite{Mateos:2000a}.

\begin{figure}
\includegraphics[width=.9\linewidth]{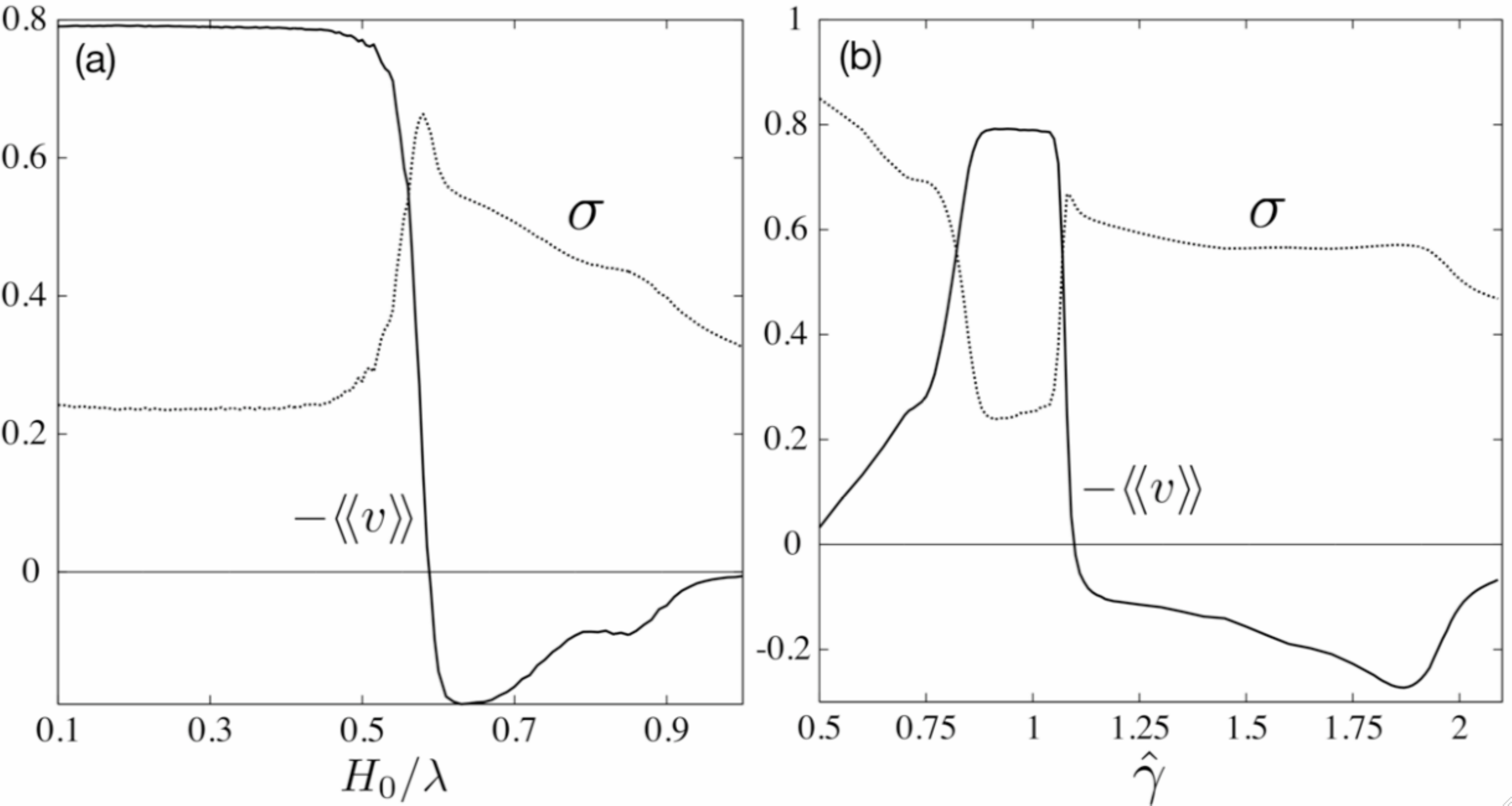}
\caption{\label{Fig:v-of-H+gamma} Mean $\langle\!\langle v
  \rangle\!\rangle$ and standard deviation $\sigma$ of the (negative)
  velocity as function of (a) the mean plate distance $H_0$ for
  $\hat\gamma=0.9$ and (b) damping $\hat\gamma$ for
  $H_0=0.1\lambda$. The other parameters are $\eta=0.65$,
  $\omega=5.0$, $\hat T=10^{-3}$.}
\end{figure}

The amplitude of the Casimir potential can be tuned by varying the
mean distance $H_0$. From Fig.~\ref{Fig:v-of-H+gamma}(a) we see that
the dynamics show a sharp transition at a critical $H_0/\lambda$ from
efficient transport with large $\langle\!\langle v \rangle\!\rangle$
and small $\sigma$ to chaotic dynamics with vanishing velocity. The
transition is accompanied by a velocity reversal and peaked velocity
fluctuations. Interestingly, below the transition $\langle\!\langle v
\rangle\!\rangle$ is almost constant independently of $H_0/\lambda$.
The observed transport behavior is also stable against a change of
effective damping $\hat\gamma$ as shown in
Fig.~\ref{Fig:v-of-H+gamma}(b). Whereas fluctuations increase with
decreasing $\hat\gamma$, there is a stable plateau of constant average
velocity across which fluctuations are diminished. In the
deterministic limit, additional plateaus with inverted and doubled
average velocity are observed by varying $\hat\gamma$ and
$\eta$. Remnants of a second plateau around $\hat\gamma=1.9$, washed
out by noise, can be seen in Fig.~\ref{Fig:v-of-H+gamma}(b).

%%%
%%% Discussion & applications
%%%

It is interesting to estimate typical velocities $v_0=\lambda/\tau$.
With the typical lengths $H_0=0.1\mu$m, $a=10$nm realized in recent
Casimir force measurements \cite{Chen:2002a} and an area mass density
of $\rho=10$g/m$^2$ for silicon plates with a thickness of a few
microns, one obtains $v_0=\sqrt{\hbar c a^2/\rho H_0^5}\approx
5.5$mm/s. The actual average velocity $v_0 \omega/2\pi$ is of the same
order for the frequencies studied above. For $\lambda=1\mu$m, the time
scale is $\tau=\lambda/v_0\approx 10^{-4}$s leading to driving
frequencies and damping rates in the kHz range for the parameters
considered here.

The results show that Casimir interactions can offer novel contact-less
translational actuation schemes for nanomechanical systems. Similar
ratchet-like effects are expected between objects of different shapes
as, e.g., periodically structured cylinders, inducing rotational
motion. The use of fluctuation forces appear also promising to move
nano-sized objects immersed in a liquid where electrostatic actuation
is not possible. Another application is the separation and detection
of particles of differing mass adsorbed to the surfaces.  For surfaces
oscillating at very high frequencies additional interesting phenomena
related to the dynamical Casimir effect occur
\cite{Golestanian:1997a}, leading to the emission of photons that
could contribute to ratchet-like effects as well.

\section{Conclusion}

There has been recent interest in applying Casimir interactions to the
design of nanomechanical devices
\cite{Chan:2001a,Chan:2001d,Buks:2001a,Buks:2001b,Buks:2002a}.  In
such devices as sensors and actuators, attractive Casimir forces can
strongly influence their function due to unwanted stiction between
small elements at nano-scale separations. But one can make also good
use of Casimir interactions in actuators where they can lead to
interesting non-linear dynamics.  Recently also repulsive Casimir
forces between bodies in a liquid, predicted some decades ago by
Lifshitz, Dzyaloshinskii and Pitaevskii for planar surfaces
\cite{Dzyaloshinskii:1961ut}, have been measured between a sphere and
a plane \cite{Munday:2009xr}, suggesting a way to suppress stiction.
Hence it is important to understand the dependence of Casimir forces
on shape and material properties beyond common approximations that apply 
only to weakly curved surfaces. This conclusion is corroborated by the
relevance of Casimir interactions to a plethora of phenomena such as
wetting, adhesion, friction, and quantum scattering of atoms from
surfaces. In this Chapter, some characteristic effects of shape and
material on Casimir interactions have been presented using the
examples of geometries that are typical to nanosystems.  Most of the
presented results could be obtained only recently by newly developed
theoretical tools that have been described here. The important study
of correlations between shape and material effects and the additional
implications of interacting fields in Casimir effects due to critical
fluctuations \cite{Hertlein:2008a} are largely unexplored. It is
expected that the recent progress on the experimental and theoretical
side will unveil novel phenomena and provide a better understanding of
fluctuation induced interactions with interesting implications for
nanosystems.

\section*{Acknowledgements}

Most of the results presented in this chapter have been obtained in
collaboration with Noah Graham, Andreas Hanke, Robert L. Jaffe, Mehran
Kardar, Sahand Jamal Rahi, and Antonello Scardicchio. Support from the
DFG grant No.~EM70/3, the Center for Theoretical Physics at MIT,
and the MIT-France Seed Fund is gratefully acknowledged.

%%%%%%%%%%%%%%%%%%%%%%%%%%%%%%%%%%%%%%

%% == Enter text of book here ==

%%%%%%%%%%%%%%%%%%%%%%%%%%%%%%%%%%%%%%
%% End of chapter
% 
% \chapappendix{Appendix title}

% \begin{chapbibliography}{10.}

% \end{chapbibliography}

%%%%%%%%%%%%%%
%% End of book:

%% Appendix
% \appendix
% \chapter{Appendix title}

%% Bibliography

%% Index
% \printindex

\end{document}